\documentclass{article}

\PassOptionsToPackage{numbers,sort&compress}{natbib}
\usepackage[preprint]{neurips_2026}

\usepackage[utf8]{inputenc}
\usepackage[T1]{fontenc}
\usepackage{hyperref}
\usepackage{url}
\usepackage{booktabs}
\usepackage{amsfonts}
\usepackage{amsmath}
\usepackage{amssymb}
\usepackage{amsthm}
\usepackage{nicefrac}
\usepackage{microtype}
\usepackage{xcolor}
\usepackage{graphicx}
\usepackage{float}
\usepackage{multirow}
\usepackage{enumitem}
\usepackage{placeins}
\usepackage{etoolbox}
\AtBeginEnvironment{tabular}{%
  \small
  \setlength{\tabcolsep}{5pt}%
}

\newcommand{\AppBone}{Appendix~\hyperref[app:b1-encoder]{B.1}}
\newcommand{\AppBtwo}{Appendix~\hyperref[app:b2-native]{B.2}}
\newcommand{\AppBthree}{Appendix~\hyperref[app:b3-kad]{B.3}}
\newcommand{\AppBfour}{Appendix~\hyperref[app:b4-supplementary]{B.4}}
\newcommand{\AppBfive}{Appendix~\hyperref[app:detailed-numeric]{B.5}}
\newcommand{\AppBsix}{Appendix~\hyperref[app:b6-toolkit]{B.6}}

\newcommand{\AppAone}{Appendix~\hyperref[app:a1-surrogate]{A.1}}
\newcommand{\AppAtwo}{Appendix~\hyperref[app:a2-theorem]{A.2}}
\newcommand{\AppAthree}{Appendix~\hyperref[app:a3-prop1]{A.3}}
\newcommand{\AppAfour}{Appendix~\hyperref[app:a4-prop2]{A.4}}

\newcommand{\ThmOne}{Theorem~\hyperlink{thm:theorem1}{\textup{1}}}
\newcommand{\PropOne}{Proposition~\hyperlink{prop:prop1}{1}}
\newcommand{\PropTwo}{Proposition~\hyperlink{prop:prop2}{2}}
\newcommand{\CorOne}{Corollary~\hyperlink{cor:corollary1}{1}}

\title{Optimal Transport Audio Distance with Learned\\Riemannian Ground Metrics}

\author{%
  Wonwoo Jeong\thanks{\texttt{jeongwonwoo@sogang.ac.kr}} \\[0.75ex]
  Sogang University \\
}

\begin{document}
\maketitle

\begin{abstract}
In audio generation evaluation, Fr\'echet Audio Distance (FAD) is a 2-Wasserstein distance with structural constraints for both primitives: the cost is a frozen embedding pullback whose invariance set hides severe artifacts, and the coupling is a Gaussian fit that dilutes rank-1 contamination relative to discrete OT. We propose Optimal Transport Audio Distance (OTAD), which corrects each primitive with one dedicated mechanism---a residual Riemannian ground-metric adapter for the cost and entropic Sinkhorn optimal transport for the coupling. Across eight encoders under a four-axis protocol, coupling-only comparisons at \(\varepsilon{=}0.05\) show that Sinkhorn's rank-1 sensitivity exceeds FAD's by a factor of \(1.9\) to \(3.6\). Furthermore, OTAD achieves a higher mean Spearman correlation with audio-quality MOS (DCASE 2023 Task~7) than baseline metrics. As an intrinsic benefit of the discrete transport plan, OTAD yields per-sample diagnostics with AUROC \(\geq 0.86\), a capability that scalar- or kernel-aggregated metrics structurally lack.
\end{abstract}

\section{Introduction}\label{sec:intro}

Text-to-audio (TTA) synthesis \cite{kreuk2023audiogen,liu2023audioldm} has matured to the point where most generated samples are perceptually adequate; user experience is now determined by sparse extreme artifacts---a misplaced click, a tonal break, a repeated motif. Evaluation metrics must therefore possess \emph{micro-level discriminability}: the ability to detect rare but consequential quality failures. Fr\'echet Audio Distance (FAD) \cite{kilgour2019fad}, the dominant audio metric, was designed for distributional comparison and is structurally insensitive to such artifacts.

We adopt a unifying viewpoint in which the FAD and Kernel Audio Distance (KAD) metrics are instances of the 2-Wasserstein distance, specified by two primitives: a \emph{cost} quantifying pairwise dissimilarity between samples, and a \emph{coupling} aggregating local matches into a global distance. FAD restricts both: its squared Euclidean cost has an invariance set onto which severe artifacts project to near-zero distance \cite{jeong2026encoderbias}, and its Gaussian coupling dilutes rank-1 contamination by a spectrum-dependent attenuation factor relative to discrete OT (\ThmOne). We propose \emph{Optimal Transport Audio Distance} (OTAD), which corrects each primitive with one dedicated mechanism.

Our contributions are threefold: \emph{(i)} a Wasserstein two-primitive lens that locates FAD's shortcomings in cost-side invariance and coupling-side rank-1 dilution (\ThmOne, \PropOne{}, and \PropTwo{}) and exposes KAD's residual cost-side restriction (Section~\ref{sec:discussion}); \emph{(ii)} OTAD, the first audio metric to correct both Wasserstein primitives---via a residual Riemannian ground-metric adapter on the cost and entropic Sinkhorn optimal transport on the coupling---yielding per-sample diagnostics that scalar- or kernel-aggregated alternatives structurally lack; \emph{(iii)} systematic empirical validation across eight encoders and a four-axis evaluation suite, providing an encoder-dependent factor decomposition and recovering the predicted rank-1 attenuation. We release the resulting metric as \texttt{otadtk}\footnote{\url{https://github.com/wonwoo-jeong/otadtk}}, serving as a drop-in replacement for existing pipelines.

\section{Related work}\label{sec:related}

The standard pipeline for audio generation evaluation relies on frozen encoder representations. While Inception Score \cite{salimans2016improved} assesses samples intrinsically via classifier entropy, FID \cite{heusel2017gans} and FAD \cite{kilgour2019fad} established the paradigm of comparing generated and reference distributions. Recently, Kernel Audio Distance (KAD) \cite{chung2025kad} relaxed FAD's Gaussian fit via an MMD kernel \cite{binkowski2018demystifying}. Although these updates primarily refine how distributions are compared, empirical critiques have probed the encoder side: \cite{gui2024adapting} compares embedding and reference-set choices for generative-music FAD across VGGish \cite{hershey2017cnn}, CLAP \cite{wu2023clap}, MERT \cite{li2024mert}, EnCodec \cite{defossez2023encodec}, and DAC \cite{kumar2023dac}, with sample-size correction and per-song diagnostics, while we previously attributed FAD's perceptual blind spots to an \emph{approximate invariance set} induced by the encoder's training task, quantified along a four-axis recall/precision/semantic/structural framework \cite{jeong2026encoderbias}; \cite{parmar2022aliased} reports analogous sensitivities in vision FID. These observations document that the encoder-induced cost is the binding constraint, yet the cost function itself is left frozen.

The methodological tools to act on this constraint come from the optimal transport literature. Beyond the Bures-Wasserstein form \cite{peyre2019computational} underlying FID/FAD, scalable couplings such as Sinkhorn divergences \cite{cuturi2013sinkhorn,feydy2019interpolating,genevay2018learning}, Wasserstein GANs \cite{arjovsky2017wasserstein}, and sliced Wasserstein distances \cite{kolouri2019generalized} relax distributional assumptions while leaving the cost as given. The decisive perspective for our setting is \emph{ground metric learning} \cite{cuturi2014ground}, which formalises learning the cost itself from data, but it has remained largely unexplored for audio evaluation. OTAD is designed precisely to bridge this gap : a learned Riemannian adapter on top of the cost-defining frozen encoder is composed with discrete entropic optimal transport on the coupling.
\section{FAD as a constrained Wasserstein distance}\label{sec:fad}

\subsection{The 2-Wasserstein distance and its two primitives}\label{sec:fad-wasserstein}

Let \(\mathcal{X}\) denote the space of audio signals. Given reference distribution \(\mu\) and test distribution \(\nu\) on \(\mathcal{X}\) equipped with a nonnegative measurable cost \(c:\mathcal{X}\times\mathcal{X}\to[0,\infty)\), the squared \(2\)-Wasserstein distance is
\[W_{2}^{2}(\mu,\nu)=\inf_{\gamma\in\Pi(\mu,\nu)}\int c(x,\tilde{x})\,d\gamma(x, \tilde{x}), \tag{1}\label{eq:1}\]
where \(\Pi(\mu,\nu)\) is the set of all joint distributions with marginals \(\mu\) and \(\nu\). This formulation involves exactly two primitives. The first is the \emph{cost function} \(c\) (Primitive 1), which quantifies pairwise dissimilarity between individual samples; every perceptual distinction the metric can detect must be encoded in \(c\). The second is the \emph{coupling measure} \(\gamma\) (Primitive 2), which specifies how mass is transported from \(\mu\) to \(\nu\); the coupling determines how distributional structure---outliers, modes, tails---contributes to the aggregate distance. Any restriction of either primitive propagates directly to \(W_{2}^{2}\): if \(c\) is blind to an artifact, no coupling can detect it; if \(\gamma\) absorbs outlier structure, no cost function can recover it. FAD restricts both, as we now formalise.

\vspace{10pt}

\textbf{Ceiling-restricted surrogates.} Two foundational inequalities, applied to FAD and KAD respectively, formalise this picture: for any distributions \(\mu,\nu\) on \(\mathbb{R}^{d}\) with finite second moments,
\[\mathrm{FAD}(\mu,\nu)\ \leq\ W_{2}^{2}(\mu,\nu),\qquad\mathrm{KAD}(\mu,\nu) \ \leq\ \mathrm{Lip}(\varphi_{k})^{2}\cdot W_{1}^{2}(\mu,\nu), \tag{2}\label{eq:2}\]
with equality in the first whenever \(\mu\) and \(\nu\) are both Gaussian, and where \(\varphi_{k}\) is the canonical feature map of the kernel \(k\) and \(\mathrm{Lip}(\varphi_{k})\) its Lipschitz constant (derivation in \AppAone). The first bound is the Gelbrich inequality \cite{dowson1982frechet,gelbrich1990formula} applied to the Gaussian projections used by FAD; the second follows from the Sriperumbudur-Sinkhorn comparison of MMD with \(W_{1}\)~\cite{feydy2019interpolating,sriperumbudur2012empirical}. Together Eq.~\eqref{eq:2} formalises our perspective: rather than recovering these OT ceilings faithfully, FAD and KAD are \emph{ceiling-restricted} surrogates that systematically attenuate distributional discrepancies. The dimension-dilution result below (\ThmOne) makes the attenuation rate explicit for FAD under rank-1 contamination; the analogous KAD signature is empirically diagnosed in Section~\ref{sec:discussion}.
\subsection{Restriction of the cost primitive: encoder-induced invariance}\label{sec:fad-cost}

A frozen encoder \(\Phi:\mathcal{X}\to\mathbb{R}^{d}\) induces the squared-Euclidean cost
\[c_{\Phi}(x,\tilde{x}):=\|\Phi(x)-\Phi(\tilde{x})\|^{2}. \tag{3}\label{eq:3}\]For tolerance \(\eta>0\), define the \emph{$\eta$-invariance set}
\[\mathcal{I}_{\Phi}(\eta):=\bigl\{\delta\in\mathcal{X}\ :\ \sup_{x\in\mathcal{X}} \|\Phi(x+\delta)-\Phi(x)\|\leq\eta\bigr\}. \tag{4}\label{eq:4}\]
By construction, \(c_{\Phi}(x,x+\delta)\leq\eta^{2}\) for every \(\delta\in\mathcal{I}_{\Phi}(\eta)\) and every \(x\in\mathcal{X}\). Therefore, if the test distribution \(\nu\) is generated by applying a perturbation \(\delta\in\mathcal{I}_{\Phi}(\eta)\) to the reference \(\mu\), the squared transport cost \(W_{2}^{2}(\mu,\nu)\) under \(c_{\Phi}\) is bounded above by \(\eta^{2}\) by definition of the infimum, regardless of the perceptual severity of the perturbation; no coupling \(\gamma\in\Pi(\mu,\nu)\) can resolve such perturbations at a finer scale than \(\eta^{2}\). The set \(\mathcal{I}_{\Phi}(\eta)\) is non-trivial whenever \(\Phi\) is trained with a non-injective objective (classification, contrastive, masked reconstruction); its empirical structure across audio encoders along recall, precision, and semantic/structural alignment axes is documented in our prior work \cite{jeong2026encoderbias}.
\subsection{Restriction of the coupling primitive: Bures-Wasserstein surrogate}\label{sec:fad-coupling}

FAD fits single Gaussians \(\mathcal{N}(\hat{\mu}_{R},\hat{\Sigma}_{R})\) and \(\mathcal{N}(\hat{\mu}_{T},\hat{\Sigma}_{T})\) to the encoder embeddings, yielding the closed-form Bures-Wasserstein distance:
\[\mathrm{FAD}=\|\hat{\mu}_{R}-\hat{\mu}_{T}\|^{2}+\mathrm{Tr}(\hat{\Sigma}_{R}+ \hat{\Sigma}_{T}-2\bigl(\hat{\Sigma}_{R}^{1/2}\hat{\Sigma}_{T}\hat{\Sigma}_{R }^{1/2}\bigr)^{1/2}\bigr). \tag{5}\label{eq:5}\]
The unique Brenier map is a global affine transformation \(T(x)=Ax+b\), which absorbs local outlier structure into the aggregate covariance. When outlier energy concentrates on a low-rank subspace, the penalty is diluted across the full \(d\)-dimensional trace:

\hypertarget{thm:theorem1}{}\textbf{Theorem 1} (Bures averaging attenuates rank-1 contamination; proof in \AppAtwo).:

\noindent\emph{Let \(\mu=\mathcal{N}(\mathbf{0},\Sigma)\) in \(\mathbb{R}^{d}\) with leading eigenpair \((\sigma_{\max}^{2},\mathbf{v}_{1})\) and trace \(T:=\mathrm{Tr}(\Sigma)\). Let \(\nu=(1-\varepsilon)\mu+\varepsilon\delta_{\mathbf{o}}\) be the rank-1 Dirac contamination at \(\mathbf{o}=L\mathbf{v}_{1}\) with outlier amplitude \(L=c_{0}\,\sigma_{\max}\) for some constant \(c_{0}\geq 1\) sufficiently large and let \(P_{n},Q_{n}\) be samples of size \(n\) drawn from \(\mu,\nu\). Let \(\mathrm{FAD}(\mu,\nu)\) denote the Bures-Wasserstein distance~\textup{(Eq.~\ref{eq:5})}. Then for \(\varepsilon\leq 1/2\) and \(n\geq n_{0}(\varepsilon,r_{\mathrm{eff}}(\Sigma))\) there exist absolute constants \(C_{1},c_{1}>0\) such that (i) \(\mathrm{FAD}(\mu,\nu)\leq\varepsilon^{2}\bigl(C_{1}L^{2}+\frac{1}{2}T\bigr)\), and (ii) \(W_{2}^{2}(P_{n},Q_{n})\geq c_{1}\,\varepsilon T\) with probability at least \(1/2\) over the empirical samples. Hence \(\mathrm{FAD}(\mu,\nu)/W_{2}^{2}(P_{n},Q_{n})=O(\varepsilon)\) with a multiplicative factor of order \(L^{2}/T=c_{0}^{2}/r_{\mathrm{eff}}(\Sigma)\), where \(r_{\mathrm{eff}}(\Sigma):=T/\sigma_{\max}^{2}\) is the spectrum's effective rank: a flat \(K\)-bounded spectrum (\(r_{\mathrm{eff}}=\Theta(d)\)) collapses the ratio to \(O(\varepsilon/d)\), while a top-eigenvalue-dominant spectrum (\(r_{\mathrm{eff}}=\Theta(1)\)) preserves \(\Theta(\varepsilon)\). Discrete OT and Sinkhorn (up to regularisation slack) attain the \(\Theta(\varepsilon T)\) scale in~\textup{(ii)}, independently of \(r_{\mathrm{eff}}\).}

By \CorOne{} (\AppAtwo), the FAD-to-\(W_{2}^{2}\) ratio is governed by the encoder spectrum's effective rank rather than by the ambient dimension alone, while discrete OT preserves a spectrum-independent contamination scale (Sinkhorn inherits this scale up to regularisation slack). The two restrictions---blind cost and dilutive coupling---together motivate the corrective design developed in Section~\ref{sec:method}.
\section{Method}\label{sec:method}

\subsection{Cost primitive: a Riemannian ground-metric adapter}\label{sec:method-cost}

Building on the ground-metric learning framework~\cite{cuturi2014ground}, we bridge deep learning and measure theory by learning a residual adapter \(g_{\theta}(z)=z+f_{\theta}(z)\)~\cite{he2016deep} that yields the corrected cost \(c_{\theta}(x,\tilde{x})=\|g_{\theta}(\Phi(x))-g_{\theta}(\Phi(\tilde{x}))\|^{2}\). This parameterization induces a linear chain of mathematical identities: the residual \(f_{\theta}\) perturbs the Euclidean identity to define a local Riemannian metric \(M_{\theta}(z)=J_{\theta}(z)^{\top}J_{\theta}(z)\) (\PropOne{}), while the Jacobian factor \(|\det J_{\theta}(z)|^{-1}\) implements the pointwise density reweighting that compensates for local volume changes under the map (\PropTwo{}). By coupling control of local distances and density, OTAD restructures the embedding space to expose rare artifacts invisible to the frozen backbone.

\hypertarget{prop:prop1}{}\textbf{Proposition 1} (Pullback metric from residual learning; proof in \AppAthree).:

\noindent\emph{For differentiable \(g_{\theta}\) with Jacobian \(J_{\theta}(z)=\partial g_{\theta}/\partial z\), first-order distances in the warped embedding satisfy \(\|g_{\theta}(z)-g_{\theta}(z^{\prime})\|^{2}=(z-z^{\prime})^{\top}M_{\theta}(z)(z-z^{\prime})+O(\|z-z^{\prime}\|^{3})\) with pullback \(M_{\theta}(z)=J_{\theta}(z)^{\top}J_{\theta}(z)\). With \(g_{\theta}(z)=z+f_{\theta}(z)\) instantiated as a two-layer bottleneck MLP (\(d\to d/4\to d\), LayerNorm, GELU), \(J_{\theta}=I+J_{f_{\theta}}\) and \(M_{\theta}=I+(J_{f_{\theta}}^{\top}+J_{f_{\theta}})+J_{f_{\theta}}^{\top}J_{f_{\theta}}\); intuitively this is a metric perturbation about the identity that keeps the backbone while adding local curvature where \(\Phi\) fails to see artifacts.}

\hypertarget{prop:prop2}{}\textbf{Proposition 2} (Jacobian volume distortion reweights probability densities; proof in \AppAfour).:

\noindent\emph{If \(g_{\theta}\) is a \(C^{1}\)-diffeomorphism and \(\mu,\nu=(g_{\theta})_{\#}\mu\) admit Lebesgue densities \(p_{\mu},p_{\nu}\), then \(p_{\nu}(g_{\theta}(z))=p_{\mu}(z)\,|\det J_{\theta}(z)|^{-1}\). Thus \(|\det J_{\theta}(z)|^{-1}\) acts as the pointwise density reweighting factor that compensates for local volume changes, redistributing probability mass according to the geometric deformation induced by \(g_{\theta}\). The adapter is not strictly invertible, but its identity-initialized residual form keeps \(|\det J_{\theta}|^{-1}\) a meaningful local signal for reshaping the embedding. For the residual map with small \(\|J_{f_{\theta}}\|_{\mathrm{op}}\), \(\det(I+J_{f_{\theta}})=1+\mathrm{tr}(J_{f_{\theta}})+O(\|J_{f_{\theta}}\|_{\mathrm{op}}^{2})\); thus a positive trace swells neighbourhoods (favouring outlier separation) and a negative trace pulls mass toward inliers.}
\subsection{Coupling primitive: entropy-regularised optimal transport}\label{sec:method-coupling}

With the cost function corrected, we replace the Gaussian coupling with entropy-regularised OT over the corrected embeddings:
\[\operatorname{OTAD}(R,T)=S_{\varepsilon}\bigl(\{g_{\theta}(\Phi(x))\}_{x\in R },\ \{g_{\theta}(\Phi(\tilde{x}))\}_{\tilde{x}\in T}\bigr), \tag{6}\label{eq:6}\]
where \(S_{\varepsilon}\) is the debiased Sinkhorn divergence \cite{feydy2019interpolating}. By Theorem~\hyperlink{thm:theorem1}{\textup{1}}\textup{(ii)}, this inherits the discrete-OT coupling scale up to Sinkhorn slack. The transport plan \(T_{ij}\) enables per-sample decomposition: the marginal cost for generated sample \(j\) is
\[c_{j}=\sum_{i}T_{ij}\cdot\|g_{\theta}(\Phi(x_{i}))-g_{\theta}(\Phi(\tilde{x}_{ j}))\|^{2}, \tag{7}\label{eq:7}\]
where high-\(c_{j}\) samples are artifact candidates. This decomposition is structurally absent from FAD's single-scalar output and is not directly provided by KAD's kernel-aggregated formulation.

\subsection{Two-track adapter training}\label{sec:method-training}

All encoders are frozen; only the lightweight adapter (\(\sim\)\(d^{2}/2\) parameters per encoder) is trained on FSD50K \cite{fonseca2022fsd50k} (200 classes, \(\sim\)37k clips). We maintain two adapter variants to serve distinct purposes.

The first variant, \(g_{\text{agnostic}}\), is trained with a metric-agnostic triplet contrastive loss \(\mathcal{L}_{\operatorname{ctr}}=\sum_{(a,p,n)}\max(0,d_{\theta}(a,p)-d_{ \theta}(a,n)+m)\), where \(d_{\theta}(a,b)=\|g_{\theta}(\Phi(a))-g_{\theta}(\Phi(b))\|^{2}\). Triplets are sampled equally from four perceptual probes---recall (same-class vs. different-class), semantic (class replacement), precision (noise degradation at varying SNR), and structural (temporal shuffle with crossfade)---ensuring the cost function is optimised without reference to any specific coupling method. This metric independence is essential for the \(2\times 2\) factor decomposition in Section~\ref{sec:exp-factor}: because \(g_{\text{agnostic}}\) treats FAD and Sinkhorn symmetrically, it provides an unbiased cost correction that enables fair separation of cost and measure effects.

The second variant, \(g_{\text{native}}\), is initialised from \(g_{\text{agnostic}}\) and fine-tuned with the Sinkhorn divergence itself: \(\mathcal{L}_{\operatorname{native}}=\mathcal{S}_{\varepsilon}(\{g_{\theta}( \Phi(x))\}_{x\in B_{R}},\{g_{\theta}(\Phi(x))\}_{\tilde{x}\in B_{T}})\). This train-eval alignment maximises the metric's correlation with perceptual quality. However, as shown in Section~\ref{sec:exp-analysis} and \AppBtwo, \(g_{\text{native}}\) reshapes the embedding space to specifically favour the Sinkhorn coupling and is therefore reserved for maximum-performance evaluation rather than for fair decomposition.
\section{Experiments}\label{sec:experiments}

\subsection{Experimental setup}\label{sec:exp-setup}

We train adapters on the FSD50K train split (200 classes, \(\sim\)37k clips) and evaluate on ESC-50 \cite{piczak2015esc50} (50 classes, 2,000 clips); no ESC-50 clips are used during training. Five primary encoders span \(d\)\(\in\)\(\{128,512,768,2048\}\)---VGGish \cite{hershey2017cnn}, EnCodec \cite{defossez2023encodec}, CLAP \cite{wu2023clap}, AudioMAE \cite{huang2022audiomae}, PANNs \cite{kong2020panns}---with OpenL3, AST, BEATs supplementing in \AppBfour{} (full specs: Table~\ref{tab:v1-encoders}, \AppBone). Each adapter is a residual MLP (\(d\to d/4\to d\), LayerNorm, GELU, dropout 0.1), trained per encoder; the metric and adapters are released as \texttt{otadtk} (\AppBsix).

Evaluation follows our four-axis protocol \cite{jeong2026encoderbias} on ESC-50 (\(R\) = all 2{,}000 clips): \emph{recall} (same-class different instances, averaged over 5 draws), \emph{semantic} (\(\varepsilon\)-fraction cross-class replacement), \emph{precision} (additive Gaussian noise at controlled SNR), and \emph{structural} (temporal segment shuffling with crossfade), each designed to stress a specific encoder weakness.
\subsection{Outlier sensitivity under rank-1 contamination}\label{sec:exp-rank1}

We begin with the cleanest possible probe of the coupling primitive: we hold the cost function fixed (no adapter, raw encoder embedding) and vary only the coupling. This isolates Primitive 2 from any cost-side effect and directly tests \ThmOne, providing the empirical baseline against which the joint cost-coupling factor decomposition in Section~\ref{sec:exp-factor} can be interpreted. We contaminate ESC-50 at rates \(\varepsilon\in\{0.005,0.01,0.02,0.05,0.1,0.15,0.2\}\) under two conditions. In the \emph{full-rank} condition, each outlier is independent Gaussian noise, spreading energy across all eigenvalues so that the covariance trace increases uniformly. In the \emph{rank-1} condition, all outliers are the same point \(\mu+5\sigma\cdot v_{1}\) along the first principal component, concentrating energy on a single eigenvalue. Four metrics are compared: FAD (Gaussian coupling), KAD (MMD kernel), Sinkhorn divergence, and Exact OT via a network simplex solver \cite{peyre2019computational}, serving as the ground-truth reference.

Under full-rank contamination, all three OT-based metrics agree to within \(0.2\%\) after \emph{self-normalisation}~\cite{jeong2026encoderbias} (each metric divided by its full-rank response at \(\varepsilon{=}0.20\)): when outlier energy is isotropic the Gaussian assumption introduces no measurable bias and the choice of coupling method is irrelevant.

\begin{figure}[t]
  \centering
  \includegraphics[width=\linewidth]{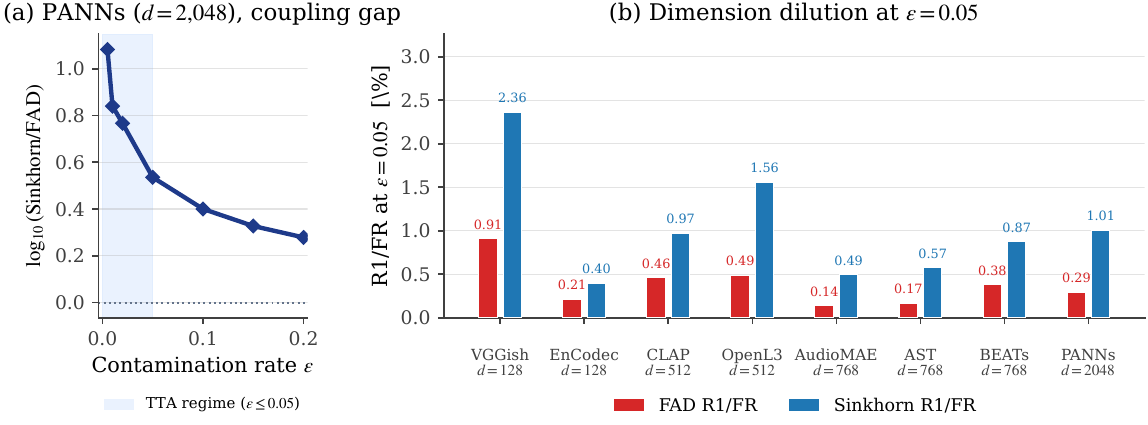}
  \caption{Rank-1 sensitivity on ESC-50 (raw embeddings; coupling-only). Panel (a): $\log_{10}(\mathrm{Sinkhorn}/\mathrm{FAD})$ on PANNs ($d{=}2{,}048$) after per-metric FR normalisation at $\varepsilon{=}0.20$; shaded band, $\varepsilon\leq 0.05$; $y{=}0$ marks parity. Panel (b): R$1/$FR (\%), rank-1 at $\varepsilon{=}0.05$ vs.\ full-rank at $\varepsilon{=}0.20$, paired FAD vs.\ Sinkhorn over eight encoders sorted by $d$; Sinkhorn dominates FAD everywhere, and the gap broadens with the effective rank of the encoder spectra (which generally scales with $d$).}
  \label{fig:v1-rank1}
\end{figure}

Rank-1 contamination tells a dramatically different story. Figure~\ref{fig:v1-rank1}(a) plots \(\log_{10}(\mathrm{Sinkhorn}/\mathrm{FAD})\) on PANNs (\(d{=}2,048\)): each metric's rank-1 response is normalised by its own full-rank baseline at \(\varepsilon{=}0.20\) before taking the ratio, and the shaded band marks the realistic TTA regime (\(\varepsilon\leq 0.05\)) where mode-collapse contamination is most plausible. In this regime FAD lies an order of magnitude below Sinkhorn, narrowing to a \(1.3{-}1.9\times\) gap only as \(\varepsilon\geq 0.15\) when both metrics partially recover. Panel (b) fixes \(\varepsilon{=}0.05\) and reports R1/FR ratios across all eight encoders: Sinkhorn dominates FAD on every encoder and the multiplicative gap broadens with embedding dimension, consistent with the spectrum-dependent FAD/\(W_{2}^{2}\) attenuation predicted by \ThmOne{} (the encoder spectra in our suite become longer-tailed at higher \(d\), lowering the \(L^{2}/T\) factor that bounds FAD's relative response), while Sinkhorn remains aligned with discrete OT up to regularisation slack. KAD fares no better here: for \(\varepsilon\leq 0.05\) its values fluctuate around zero, indistinguishable from noise, reflecting the bounded-kernel surrogate ceiling Eq.~\eqref{eq:2}. Per-encoder, per-\(\varepsilon\) values for all four metrics under both contamination conditions are tabulated in \AppBfive.

TTA mode collapse---where a model repeatedly generates the same artifact---approximates rank-1 perturbation in embedding space, precisely the regime in which FAD's Gaussian coupling is structurally attenuated while Sinkhorn retains spectrum-independent coupling sensitivity relative to discrete OT (\ThmOne). Replacing Gaussian coupling with Sinkhorn---OTAD's design choice for Primitive 2---directly closes this failure mode without invoking exact OT, which only serves as the theoretical ceiling on what any coupling-based metric can achieve.

\begin{figure}[t]
  \centering
  \includegraphics[width=0.97\linewidth]{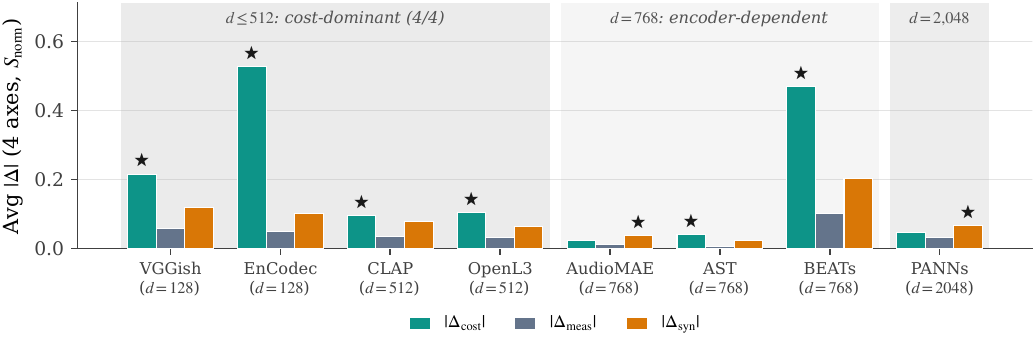}
  \caption{Average factor magnitudes $|\Delta_{\mathrm{cost}}|$, $|\Delta_{\mathrm{meas}}|$, $|\Delta_{\mathrm{syn}}|$ over the four evaluation axes for each of eight encoders, sorted by embedding dimension (light vertical shading delineates $d{\leq}512$, $d{=}768$, and $d{=}2{,}048$ bands). Cost dominates at $d{\leq}512$ (4/4 encoders); at $d{=}768$ behaviour is encoder-dependent (AudioMAE$\rightarrow$synergy, AST mixed, BEATs cost-dominant); PANNs at $d{=}2{,}048$ is synergy-dominant. Measure correction alone is never dominant.}
  \label{fig:v1-factor}
\end{figure}

\begin{table}[t]
  \caption{$2\times 2$ factor decomposition summary using $g_{\mathrm{agnostic}}$ across all eight encoders. Ranges show the magnitude of each factor across the four evaluation axes. ``Dom.'' reports the most frequent dominant factor and the number of axes (out of four) on which it dominates; ties (AST: 2C/2S) are flagged.}
  \label{tab:v1-factor-summary}
  \centering
  \begin{tabular}{@{}lccccc@{}}
    \toprule
    Encoder & \(d\) & \(|\Delta_{\rm cost}|\) range & \(|\Delta_{\rm meas}|\) range & \(|\Delta_{\rm syn}|\) range & Dom.\ (axes) \\
    \midrule
    VGGish & 128 & \textbf{0.11--0.33} & 0.02--0.09 & 0.04--0.17 & COST (3/4) \\
    EnCodec & 128 & \textbf{0.22--0.69} & 0.02--0.08 & 0.03--0.17 & COST (4/4) \\
    CLAP & 512 & \textbf{0.00--0.17} & 0.00--0.08 & 0.03--0.16 & COST (3/4) \\
    OpenL3 & 512 & \textbf{0.02--0.33} & 0.00--0.10 & 0.01--0.21 & COST (3/4) \\
    \midrule
    AudioMAE & 768 & 0.00--0.04 & 0.00--0.03 & \textbf{0.02--0.06} & SYN (3/4) \\
    AST & 768 & 0.00--0.09 & 0.00--0.02 & 0.00--0.05 & MIXED (2C/2S) \\
    BEATs & 768 & \textbf{0.36--0.59} & 0.05--0.16 & 0.10--0.32 & COST (4/4) \\
    \midrule
    PANNs & 2,048 & 0.03--0.07 & 0.02--0.06 & \textbf{0.03--0.12} & SYN (4/4) \\
    \bottomrule
  \end{tabular}
\end{table}

\subsection{Factorial decomposition of cost and coupling effects}\label{sec:exp-factor}

Section~\ref{sec:exp-rank1} fixed the cost function and showed that the coupling primitive alone produces a clear dimension-dependent signal under rank-1 contamination. We now lift the cost-fixed restriction and ask the central design question: does OTAD's improvement over FAD come primarily from the cost correction (adapter), the measure correction (Sinkhorn), or their interaction? We design a \(2\times 2\) factorial experiment with two binary factors: cost function (original \(\Phi\) vs. adapted \(g_{\theta}\circ\Phi\)) and coupling method (Gaussian vs. Sinkhorn OT). This yields four conditions: (A) FAD baseline (\(\Phi\), Gaussian), (B) measure-only (\(\Phi\), Sinkhorn), (C) cost-only (\(g_{\theta}\circ\Phi\), Gaussian), and (D) full OTAD (\(g_{\theta}\circ\Phi\), Sinkhorn). All four conditions use \(g_{\rm agnostic}\) to ensure unbiased decomposition (Section~\ref{sec:method-training}).

Because FAD and Sinkhorn produce values on different scales, direct comparison of raw numbers would be misleading. We apply a log-normalisation \(S_{\rm norm}(x)=\log(1+x)/\log(1+x_{\rm max})\) separately within each metric family (FAD for conditions A,C; Sinkhorn for B,D), mapping all values to \([0,1]\). The factor decomposition then proceeds as:
\begin{gather*}
\Delta_{\rm cost} =(C_{n}+D_{n})/2-(A_{n}+B_{n})/2, \tag{8}\label{eq:8}\\
\Delta_{\rm meas} =(B_{n}+D_{n})/2-(A_{n}+C_{n})/2, \tag{9}\label{eq:9}\\
\Delta_{\rm syn} =D_{n}-C_{n}-B_{n}+A_{n}, \tag{10}\label{eq:10}
\end{gather*}
where the sign tracks the direction of the adapter's effect on the (log-normalised) metric value; we summarise via magnitudes \(|\Delta|\) in the main text (signed per-cell values in \AppBfive).

Figure~\ref{fig:v1-factor} visualises the per-encoder factor magnitudes; Table~\ref{tab:v1-factor-summary} summarises the per-axis ranges and the dominant factor. The clearest signal is at low dimension: every \(d{\leq}512\) encoder is cost-dominant when averaging factor magnitudes across axes (Figure~\ref{fig:v1-factor}), while per-axis dominance frequencies

\noindent vary (Table~\ref{tab:v1-factor-summary}); EnCodec is the extreme case---its null-space blindness lets the adapter alone reduce FAD by up to \(\sim\)\(90\%\). At \(d{=}768\) the picture is encoder-dependent: AudioMAE shifts to synergy (3/4 axes), AST splits evenly (2C/2S), and BEATs remains strongly cost-dominant (4/4) with magnitudes rivalling EnCodec. Only at \(d{=}2{,}048\) (PANNs) does synergy dominate cleanly across all axes.

Three findings emerge. First, cost correction is the consistent lever at low dimension (13/16 cells cost-dominant at \(d{\leq}512\) in \AppBfive); the high-dimensional regime is heterogeneous (9/16 synergy-dominant cells at \(d{\geq}768\), rest cost-dominant, with BEATs the principal counter-example), pointing to architecture and pretraining as confounders we revisit in Section~\ref{sec:discussion}. Second, measure correction alone never exceeds \(|\Delta_{\mathrm{meas}}|{=}0.16\) and is the smallest factor in every encoder---replacing FAD's Gaussian with Sinkhorn, without also correcting the cost, yields negligible improvement. This shows the "Gaussian-is-the-problem" narrative, though partly true (\ThmOne), is incomplete. Third, where synergy does emerge it reveals a sequential dependence: the adapter restructures the geometry and Sinkhorn then extracts information invisible to either correction alone.
\subsection{Per-sample diagnostics}\label{sec:exp-explain}

Beyond aggregate distributional comparison, OTAD provides per-sample diagnostic information through the transport costs \(c_{j}\) defined in Eq.~\eqref{eq:7}. We validate this capability through controlled contamination experiments on ESC-50; alignment with human judgement is then taken up in Section~\ref{sec:exp-mos}.

Starting from the clean reference (\(N{=}2{,}000\)), we contaminate an \(\varepsilon\)-fraction of samples with three types of artifacts: Gaussian noise, cross-class replacement (real audio from wrong categories), and silence insertion. For each contaminated distribution, we compute \(c_{j}\) for every sample and evaluate whether contaminated samples receive systematically higher costs using AUROC and the separation ratio (mean contaminated cost / mean clean cost).

\begin{figure}[t]
  \centering
  \includegraphics[width=\linewidth]{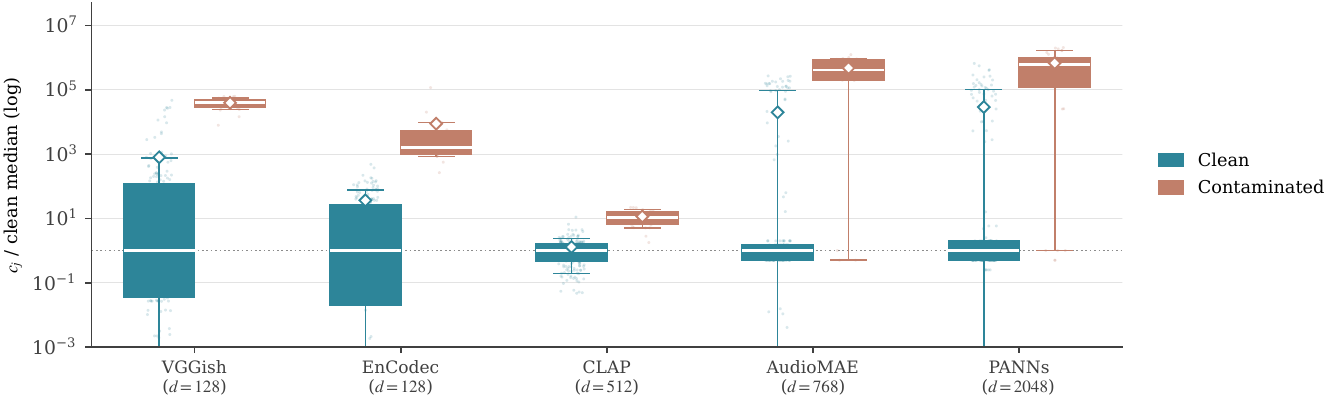}
  \caption{Per-sample transport costs $c_j$ on ESC-50 under cross-class contamination ($\varepsilon{=}0.05$), normalised to the clean median per encoder (AUROC and mean separation ratios are tabulated in \protect\AppBfive).}
  \label{fig:v1-hist}
\end{figure}

At \(\varepsilon{=}0.05\), all five encoders achieve perfect detection on the Gaussian-noise and silence settings (\(\mathrm{AUROC}{=}1.0\)), with separation ratios spanning up to five orders of magnitude across encoders. Cross-class replacement is the most challenging case because the contaminating samples are real, well-formed audio that differs from the reference only in semantic content; even here, AUROC remains in \([0.86,1.00]\) across encoders. Figure~\ref{fig:v1-hist} visualises the underlying \(c_{j}\) distributions per encoder, normalised so the clean median sits at \(1\): the contaminated boxes shift consistently above the clean reference on the logarithmic scale, with the per-encoder AUROC and separation ratio annotated. These results establish that OTAD can answer not just "how different is this distribution?" but "which samples are problematic and how problematic are they?"---a diagnostic capability structurally impossible under FAD's single-scalar formulation; full per-encoder, per-contamination metrics are catalogued in \AppBfive{} (bundled with \texttt{otadtk}).

\begin{figure}[t]
  \centering
  \includegraphics[width=0.88\linewidth]{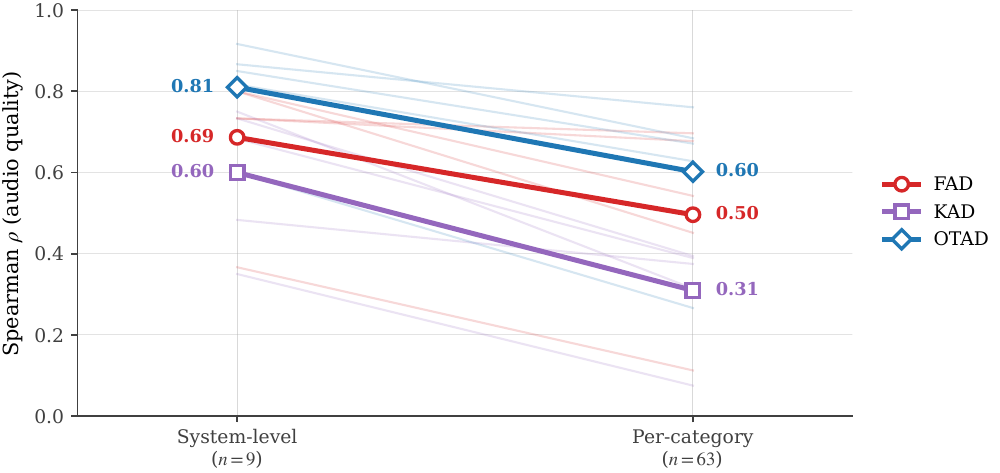}
  \caption{Mean Spearman correlation between distributional metrics and human MOS on DCASE 2023 Task 7 (audio-quality axis), averaged over the five perceptually competent encoders (faint background lines show individual encoders; EnCodec is excluded from the means). OTAD outperforms FAD and KAD at both granularities; the margin widens at per-category granularity, where KAD's correlation collapses below FAD.}
  \label{fig:v1-mos}
\end{figure}

\subsection{Alignment with human MOS}\label{sec:exp-mos}

Sections~\ref{sec:exp-rank1}--\ref{sec:exp-explain} establish OTAD's correctness on controlled perturbations. We now close the loop with an external ground truth: human Mean Opinion Scores on the DCASE 2023 Task 7 Foley Sound Synthesis benchmark \cite{choi2023dcase}, the same evaluation suite used to validate KAD \cite{chung2025kad}. The benchmark releases community-sourced MOS along two axes (audio quality and category fit) for \(9\) generative systems---\(4\) Track A entries, \(4\) Track B entries, and the official Task 7 baseline---across \(7\) Foley categories (dog bark, footstep, gunshot, keyboard, motor vehicle, rain, sneeze/cough), with \(100\) generated clips per system-category cell. For each cell we compute FAD, KAD, and OTAD against \(100\) held-out reference clips from the same category, then correlate the resulting metric values with the published MOS at two granularities: \emph{system-level} (\(n{=}9\), averaging the metric across categories within each system) and \emph{per-category} (\(n{=}63\), treating each system-category cell as an observation).

Figure~\ref{fig:v1-mos} traces how each metric's mean Spearman correlation with audio-quality MOS evolves from coarse system-level ranking (\(n{=}9\)) to fine per-category granularity (\(n{=}63\)), averaged over the five perceptually competent encoders (VGGish, PANNs-CNN14, PANNs-WGLM, CLAP, AudioMAE; EnCodec discussed separately below); KAD uses the eval-set median pairwise distance as bandwidth, matching the reference \texttt{kadtk} implementation \cite{chung2025kad}. OTAD with \(g_{\mathrm{agnostic}}\) leads at both granularities, with the lead growing as the granularity tightens: at system-level it lifts mean \(\bar{\rho}\) over both FAD and KAD by roughly \(0.1{-}0.2\), and at per-category granularity the gap widens because KAD's correlation collapses below FAD while OTAD remains stable. This collapse is the empirical signature of KAD's cost-side limitation predicted by our two-primitive analysis (Section~\ref{sec:fad-cost}): at sub-system granularity the implicit RBF cost cannot resolve fine perceptual differences within \(\mathcal{I}_{\Phi}\), so the coupling-side improvement alone fails to translate into perceptual alignment. PANNs-WGLM, the variant on which \cite{chung2025kad} reports KAD's strongest results, is the closest stress test: KAD narrows its gap to FAD most at system level (\(0.73\) vs. \(0.80\)) but still drops below FAD per-category, while OTAD remains best on both (\AppBthree). EnCodec is the lone exception: every metric---including OTAD---yields weak or negative MOS correlation (\(\rho\in[-0.21,0.07]\)), reflecting its compression-oriented training \cite{gui2024adapting} that no distance can recover, which we report as a faithful encoder-side limitation.

\subsection{Analysis of Design and Efficiency}\label{sec:exp-analysis}

The two-track training strategy is validated by comparing \(g_{\mathrm{agnostic}}\) and \(g_{\mathrm{native}}\) under the same \(2\times 2\) decomposition (full results in \AppBtwo). With \(g_{\mathrm{native}}\), every encoder switches to cost dominance and the Sinkhorn divergence collapses far more than FAD does, confirming that \(g_{\mathrm{native}}\) reshapes the embedding to minimise the Sinkhorn coupling and is therefore unsuited for fair \(2\times 2\) separation. The Sinkhorn regularisation \(\varepsilon_{\mathrm{reg}}\) has a sweet spot in \([0.05, 0.10]\) (sweep details in \AppBfive); below this range the solver iterates without gain, whereas above \(0.5\) entropy overwhelms cost information. Counterintuitively, for \(N \leq 2{,}000\), Sinkhorn is an order of magnitude faster than FAD on a standard GPU (FAD is CPU-bound by \texttt{sqrtm}), with mini-batch or sliced OT extending this scalability to larger datasets.
\section{Discussion}\label{sec:discussion}

The same decomposition that pins down FAD's failures (Section~\ref{sec:fad}) places KAD \cite{chung2025kad} between FAD and OTAD in our two-primitive lens: replacing the Gaussian coupling with an MMD kernel removes the parametric coupling assumption (KAD is the high-regularisation limit of entropic OT \cite{feydy2019interpolating,genevay2018learning}), but the bounded RBF kernel inherits the encoder invariance set \(\mathcal{I}_{\Phi}\) --- for nearby points its induced RKHS distance behaves like \(\|\Phi(x)-\Phi(\tilde{x})\|/\sigma\), yet saturates at a finite ceiling for distant points --- producing a noise floor that prevents matching discrete-OT sensitivity under rank-1 contamination. The empirical signatures are the rank-1 noise floor of Section~\ref{sec:exp-rank1} and the per-category MOS collapse of Section~\ref{sec:exp-mos}. OTAD therefore corrects the primitive that KAD leaves untouched while subsuming the coupling-side benefit of MMD as a regularisation limit.

\section{Limitations and future work}\label{sec:limitations}

Our human-MOS validation rests on a single listening test (DCASE 2023 Task 7 Foley, \(n{=}9\) systems and \(n{=}63\) system-category cells) and does not include music or speech generative benchmarks; the magnitude of OTAD's perceptual gain in those domains has not been measured, and the small system count bounds the statistical power available to separate metric variants. All adapters are trained on FSD50K (200 environmental-sound classes), which biases the cost-side correction towards general audio, and we do not provide cross-domain evidence that the same adapter generalises---or that re-trained adapters preserve their cost-side advantage---to music, speech, or specialised audio domains. Likewise, our encoder coverage spans eight general-audio backbones but excludes music-domain backbones such as MERT and DAC \cite{kumar2023dac,li2024mert}; whether OTAD retains its cost-side advantage on music-specialised representations is open. 

The residual adapter cannot recover information already discarded by the encoder's null space: EnCodec is the empirical witness, with MOS correlation remaining weak across both axes (\(\rho\in[-0.21,0.07]\)) under every metric including OTAD, confirming that distance-side correction has fundamental ceilings dictated by the upstream representation. On the coupling side, naive entropic OT becomes memory-bound beyond \({\sim}5{,}000\) samples, so mini-batch, sliced, or low-rank approximations are needed at larger scales. \ThmOne{} also formalises only the rank-1 stress test: the FAD/\(W_{2}^{2}\) ratio is bounded by an effective-rank factor \(L^{2}/T=c_{0}^{2}/r_{\mathrm{eff}}(\Sigma)\) rather than ambient dimension alone, so the empirical \(1/d\)-like attenuation in Figure~\ref{fig:v1-rank1} is mediated by encoder spectra and may not extrapolate to backbones with markedly different effective ranks; multi-mode or correlated artifacts likewise lie outside this model.

Future work should extend the MOS validation to music and speech domains, and to music-specialised encoders such as MERT and DAC, to test the generality of the cost-side advantage, while sequence-level OT models and scalable approximations are needed to handle correlated artifacts and larger evaluation scales.

\section{Conclusion}\label{sec:conclusion}

We introduced a two-primitive lens for audio evaluation, showing that FAD suffers from both an encoder-invariant cost that blinds it to artifacts and a Gaussian coupling that dilutes rank-1 contamination by a spectrum-dependent factor (Theorem~\hyperlink{thm:theorem1}{\textup{1}}). KAD relaxes the coupling but inherits the same frozen cost. OTAD corrects each primitive directly: a residual Riemannian adapter learns a local cost that escapes encoder invariances, while Sinkhorn entropic optimal transport preserves the spectrum-independent sensitivity of discrete OT. 

This dual correction recovers discrete-OT-grade rank-1 sensitivity, supplies per-sample diagnostics (\(\mathrm{AUROC}\geq 0.86\)) that scalar- and kernel-aggregated metrics structurally lack, and achieves the highest MOS correlation on DCASE 2023 Foley --- at fine granularity where KAD collapses below FAD, OTAD remains stable.
A factorial decomposition across eight encoders shows that cost correction is the dominant lever at low dimensions, that measure correction alone is never sufficient, and that a learned synergy emerges at higher dimensions.
The two-primitive lens and the residual adapter design together provide a modality-independent blueprint for any FID-style evaluation. For audio, the released \texttt{otadtk} toolkit serves as a drop-in replacement.

\clearpage
\FloatBarrier

\appendix

\section{Proofs}\label{app:proofs}

\subsection{Derivation of the ceiling-restricted surrogate bounds (Eq.~\eqref{eq:2})}\label{app:a1-surrogate}

Both inequalities express the same principle: replacing unrestricted OT objects by surrogates---Gaussian summaries for FAD or kernel-induced witnesses for KAD---contracts the induced discrepancy scores relative to \(W_{2}^{2}\) and \(W_{1}^{2}\), as summarised by Eq.~\eqref{eq:2}.

\textbf{The Bures-Wasserstein bound:} \(\mathrm{FAD}\leq W_{2}^{2}\). Let \(\mu_{G},\nu_{G}\) denote the Gaussian projections of \(\mu,\nu\) matching the mean \(m_{\mu},m_{\nu}\) and covariance \(\Sigma_{\mu},\Sigma_{\nu}\), so by definition \(\mathrm{FAD}(\mu,\nu)=W_{2}^{2}(\mu_{G},\nu_{G})\). For any coupling \(\gamma\in\Pi(\mu,\nu)\), the squared transport cost decomposes as
\[\int\|x-y\|^{2}\,d\gamma(x,y)=\|m_{\mu}-m_{\nu}\|^{2}+\mathrm{Tr}(\Sigma_{\mu} )+\mathrm{Tr}(\Sigma_{\nu})-2\ \mathrm{Tr}\bigl(\Sigma_{XY}^{\,\gamma}\bigr), \tag{11}\label{eq:11}\]
where \(\Sigma_{XY}^{\,\gamma}=\mathbb{E}_{\gamma}\bigl[(X-m_{\mu})(Y-m_{\nu})^{\top} \bigr]\) is the cross-covariance induced by \(\gamma\). By Gelbrich's theorem \cite{gelbrich1990formula}, the trace \(\mathrm{Tr}(\Sigma_{XY}^{\,\gamma})\) is bounded above, over all couplings with marginal covariances \((\Sigma_{\mu},\Sigma_{\nu})\), by \(\mathrm{Tr}\bigl((\Sigma_{\mu}^{1/2}\Sigma_{\nu}\Sigma_{\mu}^{1/2})^{1/2} \bigr)\), and this upper bound is attained by the Gaussian Bures coupling of \(\mu_{G},\nu_{G}\). Substituting into Eq.~\eqref{eq:11} and taking the infimum over \(\gamma\in\Pi(\mu,\nu)\),
\[W_{2}^{2}(\mu,\nu)\ \geq\ \|m_{\mu}-m_{\nu}\|^{2}+\mathrm{Tr}\bigl(\Sigma_{\mu} +\Sigma_{\nu}-2(\Sigma_{\mu}^{1/2}\Sigma_{\nu}\Sigma_{\mu}^{1/2})^{1/2}\bigr). \tag{12}\label{eq:12}\]
Equality requires the optimal coupling \(\gamma^{*}\) to attain the cross-covariance bound, which by Gelbrich's theorem occurs iff the marginals \(\mu\) and \(\nu\) are Gaussian (in which case the optimal coupling is jointly Gaussian).

\textbf{The MMD-\(W_{1}\) bound:} \(\mathrm{KAD}\leq\mathrm{Lip}(\varphi_{k})^{2}\cdot W_{1}^{2}\). Let \(\varphi_{k}:\mathbb{R}^{d}\to H_{k}\) be the canonical feature map associated with kernel \(k\), with Lipschitz constant \(L_{k}=\sup_{x\neq x^{\prime}}\|\varphi_{k}(x)-\varphi_{k}(x^{\prime})\|_{H_{k} }/\|x-x^{\prime}\|\). By Riesz representation,
\[\mathrm{MMD}(\mu,\nu)=\bigl\|\mathbb{E}_{\mu}\varphi_{k}-\mathbb{E}_{\nu} \varphi_{k}\bigr\|_{H_{k}}=\sup_{g\in H_{k},\,\|g\|_{H_{k}}\leq 1}\int \langle g,\varphi_{k}(x)\rangle_{H_{k}}\,d(\mu-\nu)(x). \tag{13}\label{eq:13}\]
For each unit-norm \(g\), the function \(f_{g}(x)=\langle g,\varphi_{k}(x)\rangle_{H_{k}}\) has Lipschitz constant \(\|f_{g}\|_{\mathrm{Lip}}\leq\|g\|_{H_{k}}\cdot L_{k}\leq L_{k}\) by Cauchy-Schwarz, hence belongs to the rescaled \(L_{k}\)-Lipschitz ball. Combining Eq.~\eqref{eq:13} with the Kantorovich-Rubinstein duality \(W_{1}(\mu,\nu)=\sup_{\|f\|_{\mathrm{Lip}}\leq 1}\int f\,d(\mu-\nu)\),
\[\mathrm{MMD}(\mu,\nu)\ \leq\ L_{k}\cdot\sup_{\|f\|_{\mathrm{Lip}}\leq 1}\int f\,d( \mu-\nu)\ =\ L_{k}\cdot W_{1}(\mu,\nu), \tag{14}\label{eq:14}\]
so \(\mathrm{KAD}(\mu,\nu)=\mathrm{MMD}^{2}(\mu,\nu)\leq L_{k}^{2}\cdot W_{1}^{2}( \mu,\nu)\)~\cite{feydy2019interpolating,sriperumbudur2012empirical}. For Gaussian RBF kernels \(k_{\sigma}(x,x^{\prime})=\exp(-\|x-x^{\prime}\|^{2}/2\sigma^{2})\), \(L_{k}=1/\sigma\), giving the bandwidth-dependent contraction \(\mathrm{MMD}\leq W_{1}/\sigma\) that underlies the noise floor reported in Section~\ref{sec:exp-rank1}.

\subsection{Proof of Theorem 1 (Bures averaging attenuation)}\label{app:a2-theorem}

Let \(\mu=\mathcal{N}(\mathbf{0},\Sigma)\) denote the reference in \(\mathbb{R}^{d}\), with eigenpairs \(\{(\sigma_{i}^{2},\mathbf{v}_{i})\}_{i=1}^{d}\) ordered \(\sigma_{1}^{2}\geq\cdots\geq\sigma_{d}^{2}>0\) and trace \(T:=\mathrm{Tr}(\Sigma)=\sum_{i}\sigma_{i}^{2}\). Write \(\sigma_{\max}:=\sigma_{1}\), \(\mathbf{o}=L\,\mathbf{v}_{1}\) with \(L=c_{0}\,\sigma_{\max}\), \(c_{0}\geq 1\), and \(r_{\mathrm{eff}}(\Sigma):=T/\sigma_{\max}^{2}\in[1,d]\). The contaminated mixture \(\nu=(1-\varepsilon)\mu+\varepsilon\delta_{\mathbf{o}}\) has moments
\[\tilde{\mathbf{\mu}}=\varepsilon L\,\mathbf{v}_{1},\qquad\qquad\tilde{\Sigma}=(1- \varepsilon)\Sigma+\varepsilon^{\prime}\,L^{2}\,\mathbf{v}_{1}\mathbf{v}_{1}^{ \top},\qquad\qquad\varepsilon^{\prime}:=\varepsilon(1-\varepsilon), \tag{15}\label{eq:15}\]
where the covariance follows from \(\mathbb{E}_{\nu}[XX^{\top}]=(1-\varepsilon)\Sigma+\varepsilon L^{2}\mathbf{v}_{1 }\mathbf{v}_{1}^{\top}\) and subtracting the rank-1 outer product \(\tilde{\mathbf{\mu}}\tilde{\mathbf{\mu}}^{\top}=\varepsilon^{2}L^{2}\mathbf{v}_{1} \mathbf{v}_{1}^{\top}\). The leading eigenvalue becomes \(\tilde{\sigma}_{1}^{2}=(1-\varepsilon)\sigma_{1}^{2}+\varepsilon^{\prime}L^{2}\), while every off-leading eigenvalue contracts to \(\tilde{\sigma}_{i}^{2}=(1-\varepsilon)\sigma_{i}^{2}\) for \(i\geq 2\).

Proof of \ThmOne\textup{(i)}.:

\noindent The squared Bures-Wasserstein distance between \(\mu\) and the Gaussian fit of \(\nu\) decomposes as
\[\mathrm{FAD}=\|\tilde{\mathbf{\mu}}\|^{2}+\sum_{i=1}^{d}\bigl{(}\sigma_{i}-\tilde{ \sigma}_{i}\bigr{)}^{2}=\varepsilon^{2}L^{2}+\bigl{(}\sigma_{1}-\sqrt{(1- \varepsilon)\sigma_{1}^{2}+\varepsilon^{\prime}L^{2}}\bigr{)}^{2}+\bigl{(}1- \sqrt{1-\varepsilon}\bigr{)}^{2}\sum_{i=2}^{d}\sigma_{i}^{2}. \tag{16}\label{eq:16}\]

\clearpage
We bound each term in turn.

\emph{Off-leading contraction.} The identity \(1-\sqrt{1-\varepsilon}=\varepsilon/(1+\sqrt{1-\varepsilon})\) and \(\varepsilon\leq 1/2\) give \((1-\sqrt{1-\varepsilon})^{2}\leq\varepsilon^{2}/(1+\sqrt{1/2})^{2}\leq \varepsilon^{2}/2\), so
\[\bigl(1-\sqrt{1-\varepsilon}\bigr)^{2}\!\sum_{i=2}^{d}\sigma_{i}^{2}\;\leq\; \frac{1}{2}\,\varepsilon^{2}\,(T-\sigma_{\max}^{2})\;\leq\;\frac{1}{2}\, \varepsilon^{2}\,T. \tag{17}\label{eq:17}\]
\emph{Leading direction.} Splitting \(\sigma_{1}-\sqrt{(1-\varepsilon)\sigma_{1}^{2}+\varepsilon^{\prime}L^{2}}=( \sigma_{1}-\sqrt{(1-\varepsilon)}\sigma_{1})+(\sqrt{(1-\varepsilon)}\sigma_{1 }-\sqrt{(1-\varepsilon)\sigma_{1}^{2}+\varepsilon^{\prime}L^{2}})\) and using \((a+b)^{2}\leq 2a^{2}+2b^{2}\) together with \(\sqrt{x+y}-\sqrt{x}\leq y/(2\sqrt{x})\):
\begin{equation}
\begin{aligned}
\bigl(\sigma_{1}-\tilde{\sigma}_{1}\bigr)^{2}
&\leq 2\bigl(1-\sqrt{1-\varepsilon}\bigr)^{2}\sigma_{1}^{2}
+2\,\frac{(\varepsilon^{\prime}L^{2})^{2}}{4(1-\varepsilon)\sigma_{1}^{2}}
\\
&\leq\varepsilon^{2}\sigma_{\max}^{2}+\frac{\varepsilon^{2}L^{4}}{2(1-\varepsilon)\sigma_{\max}^{2}}
\\
&\leq\varepsilon^{2}\sigma_{\max}^{2}+\varepsilon^{2}c_{0}^{2}\,L^{2}.
\end{aligned}
\tag{18}\label{eq:18}
\end{equation}
where the last step uses \(L^{2}=c_{0}^{2}\sigma_{\max}^{2}\) and \(\varepsilon\leq 1/2\). Since \(\sigma_{\max}^{2}=L^{2}/c_{0}^{2}\leq L^{2}\) this contributes \(\bigl(\sigma_{1}-\tilde{\sigma}_{1}\bigr)^{2}\leq\varepsilon^{2}(1+c_{0}^{2} )L^{2}\).

\emph{Combining.} Substituting Eq.~\eqref{eq:17}--Eq.~\eqref{eq:18} into Eq.~\eqref{eq:16},
\[\mathrm{FAD}\;\leq\;\varepsilon^{2}\bigl(C_{1}L^{2}+\frac{1}{2}T\bigr), \qquad C_{1}:=2+c_{0}^{2}, \tag{19}\label{eq:19}\]
proving~\hyperlink{thm:theorem1}{\textup{(i)}}. Equivalently, using \(T=r_{\mathrm{eff}}\,\sigma_{\max}^{2}=(r_{\mathrm{eff}}/c_{0}^{2})\,L^{2}\), the bound rewrites as \(\mathrm{FAD}\leq\varepsilon^{2}L^{2}\bigl(C_{1}+r_{\mathrm{eff}}/(2c_{0}^{2} )\bigr)\), exhibiting the explicit dependence on the spectrum's effective rank.\hfill$\square$

\emph{Proof of \ThmOne\textup{(ii)}.} Consider empirical distributions \(P_{n}=\frac{1}{n}\sum_{i=1}^{n}\delta_{x_{i}}\) with \(x_{i}\stackrel{{\mathrm{iid}}}{{\sim}}\mu\), and \(Q_{n}=\frac{1}{n}\sum_{j=1}^{n}\delta_{y_{j}}\) where \(\lceil\varepsilon n\rceil\) of the \(y_{j}\) are set to \(\mathbf{o}\) (the remainder are iid \(\mu\)-samples; their precise values do not enter the lower bound below). Define squared distances \(D_{i}=\|x_{i}-\mathbf{o}\|^{2}\) for \(i=1,\ldots,n\), and let \(D_{(1)}\leq D_{(2)}\leq\cdots\leq D_{(n)}\) be their order statistics.

Any feasible coupling \(\gamma\in\Pi(P_{n},Q_{n})\) must assign at least \(\lceil\varepsilon n\rceil\) units of mass from the reference to the outlier location. The minimal possible cost is obtained by matching the \(\lceil\varepsilon n\rceil\) closest reference points to the outlier. Hence
\[W_{2}^{2}(P_{n},Q_{n})\;\geq\;\frac{1}{n}\sum_{k=1}^{\lceil\varepsilon n\rceil}D_{(k)}. \tag{20}\label{eq:20}\]
We now lower-bound this order-statistic sum. Set a threshold \(\tau=T/2\). For \(x\sim\mathcal{N}(\mathbf{0},\Sigma)\), the squared distance \(D=\|x-\mathbf{o}\|^{2}\) is a non-central \(\chi^{2}\)-type quadratic form with first two moments bounded as follows:
\[\mathbb{E}\bigl[\|x-\mathbf{o}\|^{2}\bigr]=\mathrm{Tr}(\Sigma)-2L\,\mathbb{E }\langle x,\mathbf{v}_{1}\rangle+L^{2}=\mathrm{Tr}(\Sigma)+L^{2}\;\geq\; \mathrm{Tr}(\Sigma). \tag{21}\label{eq:21}\]
Expanding \(x=\sum_{i}\xi_{i}\sigma_{i}\mathbf{v}_{i}\) with \(\xi_{i}\stackrel{{\mathrm{iid}}}{{\sim}}\mathcal{N}(0,1)\), the quantities \(\|x\|^{2}=\sum_{i}\xi_{i}^{2}\sigma_{i}^{2}\) and \(\langle x,\mathbf{v}_{1}\rangle=\xi_{1}\sigma_{1}\) are uncorrelated (the latter is odd in \(\xi_{1}\), the former invariant), so
\begin{equation}
\begin{aligned}
\mathrm{Var}\bigl[\|x-\mathbf{o}\|^{2}\bigr]
&=\mathrm{Var}\bigl[\|x\|^{2}\bigr]+4L^{2}\,\mathrm{Var}\bigl[\langle x,\mathbf{v}_{1}\rangle\bigr]
\\
&=2\sum_{i=1}^{d}\sigma_{i}^{4}+4L^{2}\sigma_{1}^{2}
\leq 2\sigma_{\max}^{2}\,T+4c_{0}^{2}\sigma_{\max}^{4}
\\
&=2\bigl(1+2c_{0}^{2}/r_{\mathrm{eff}}\bigr)\,\sigma_{\max}^{2}\,T
\leq C_{V}\,\sigma_{\max}^{2}\,T.
\end{aligned}
\tag{22}\label{eq:22}
\end{equation}
using \(\sigma_{i}^{2}\leq\sigma_{\max}^{2}\), \(L^{2}=c_{0}^{2}\sigma_{\max}^{2}\), \(r_{\mathrm{eff}}\geq 1\), and \(C_{V}:=2(1+2c_{0}^{2})\). By standard concentration for quadratic forms of Gaussian vectors (e.g., a sub-exponential tail via Hanson--Wright, or a one-sided Chernoff bound for non-central \(\chi^{2}\)), there exists an absolute constant \(c>0\) such that, with \(\tau=T/2\),
\[\mathbb{P}(D\leq\tau)\;\leq\;\exp\!\Bigl(-c\,\frac{(\tau+L^{2})^{2}}{\sigma_{\max}^{2}\,T}\Bigr)\;\leq\;\exp\!\Bigl(-c\,\frac{L^{4}}{\sigma_{\max}^{2}\,T}\Bigr)\;=\;\exp\!\Bigl(-c\,c_{0}^{4}\,\frac{\sigma_{\max}^{2}}{T}\Bigr). \tag{23}\label{eq:23}\]
Taking \(c_{0}\) sufficiently large as allowed in the theorem statement, we may assume \(p:=\mathbb{P}(D\leq\tau)=\mathbb{P}_{x\sim\mu}(\|x-\mathbf{o}\|^{2}\leq T/2)\leq\varepsilon/4\) (Eq.~\eqref{eq:23} gives \(\mathbb{P}(D\leq\tau)\leq\exp(-c\,c_{0}^{4}\,\sigma_{\max}^{2}/T)\), which drives this tail below \(\varepsilon/4\) for \(c_{0}\) large enough).

Let \(K=\sum_{i=1}^{n}\mathbf{1}_{\{D_{i}\leq\tau\}}\) count reference points within squared distance \(\tau\) of \(\mathbf{o}\). Then \(\mathbb{E}[K]=np\leq n\varepsilon/4\). By Markov's inequality,
\[\mathbb{P}\!\Bigl(K\geq\frac{1}{2}\,\varepsilon n\Bigr)\;\leq\;\frac{\mathbb{E}[K]}{\frac{1}{2}\,\varepsilon n}\;\leq\;\frac{n\varepsilon/4}{\varepsilon n/2}\;=\;\frac{1}{2}.\]
Hence, with probability at least \(1/2\), we have \(K<\tfrac{1}{2}\varepsilon n\). On this event, at most \(\tfrac{1}{2}\varepsilon n\) of the distances \(\{D_{i}\}\) are less than \(T/2\), so among the \(\lceil\varepsilon n\rceil\) smallest distances at least \(\lceil\varepsilon n\rceil-K\geq\varepsilon n/2\) exceed \(T/2=\tau\). Therefore,
\[\frac{1}{n}\sum_{k=1}^{\lceil\varepsilon n\rceil}D_{(k)}\;\geq\;\frac{1}{n}\cdot\frac{\varepsilon n}{2}\cdot\frac{T}{2}\;=\;\frac{\varepsilon T}{4}.\]
Substituting into Eq.~\eqref{eq:20},
\[W_{2}^{2}(P_{n},Q_{n})\;\geq\;\frac{\varepsilon}{4}\,T\;=\;c_{1}\,\varepsilon\,T, \qquad c_{1}:=\frac{1}{4}, \tag{24}\label{eq:24}\]
with probability at least \(1/2\), proving~\hyperlink{thm:theorem1}{\textup{(ii)}}. The debiased Sinkhorn divergence \(S_{\varepsilon_{\mathrm{reg}}}\) inherits the same lower bound up to \(O(\varepsilon_{\mathrm{reg}})\) slack via the OT--Sinkhorn comparison of \cite{feydy2019interpolating}.\hfill$\square$

\hypertarget{cor:corollary1}{}\textbf{Corollary 1}.: \emph{Combining parts \textup{(i)} and \textup{(ii)} of \ThmOne{} and using \(L^{2}/T=c_{0}^{2}/r_{\rm eff}(\Sigma)\),}
\[\frac{\operatorname{FAD}(\mu,\nu)}{W_{2}^{2}(P_{n},Q_{n})}\ \leq\ \frac{ \varepsilon^{2}\bigl(C_{1}\,L^{2}+\frac{1}{2}T\bigr)}{c_{1}\,\varepsilon\,T} \ =\ \frac{C_{1}}{c_{1}}\cdot\frac{\varepsilon\,c_{0}^{2}}{r_{\rm eff}(\Sigma)}\ +\ \frac{ \varepsilon}{2c_{1}}\ =\ O\bigl(\varepsilon\cdot(1+1/r_{\rm eff})\bigr). \tag{25}\label{eq:25}\]
\emph{A flat \(K\)-bounded spectrum gives \(r_{\rm eff}\geq d/K\) and hence the contracted ratio \(O(\varepsilon/d)+O(\varepsilon)\), while a top-eigenvalue-dominant spectrum (\(r_{\rm eff}=O(1)\)) yields the dimension-independent \(O(\varepsilon)\) bound that tracks the per-sample \(\Theta(\varepsilon T)\) scale of discrete OT up to a constant slack.}

\subsection{Proof of Proposition 1 (pullback metric from residual learning)}\label{app:a3-prop1}

Let \(g_{\theta}:\mathbb{R}^{d}\to\mathbb{R}^{d}\) be a differentiable adapter with Jacobian \(J_{\theta}(z)=\partial g_{\theta}/\partial z\in\mathbb{R}^{d\times d}\). Define the pullback metric tensor \(M_{\theta}(z)=J_{\theta}(z)^{\top}J_{\theta}(z)\). We show that the squared Euclidean distance in the transformed space reduces to a local Mahalanobis distance under \(M_{\theta}\), and that the residual parameterisation implements a \emph{metric perturbation}---an additive deformation of the identity inner product induced by \(f_{\theta}\).

Proof.:

\noindent Let \(\delta=z^{\prime}-z\). Taylor-expand \(g_{\theta}\) around \(z\):
\[g_{\theta}(z^{\prime})=g_{\theta}(z)+J_{\theta}(z)\,\delta+O(\|\delta\|^{2}). \tag{26}\label{eq:26}\]
Subtracting and taking the squared norm:
\begin{equation}
\begin{aligned}
\|g_{\theta}(z^{\prime})-g_{\theta}(z)\|^{2}
&=\|J_{\theta}(z)\,\delta+O(\|\delta\|^{2})\|^{2}
\\
&=\delta^{\top}J_{\theta}(z)^{\top}J_{\theta}(z)\,\delta+O(\|\delta\|^{3})
\\
&=\delta^{\top}M_{\theta}(z)\,\delta+O(\|\delta\|^{3}).
\end{aligned}
\tag{27}\label{eq:27}
\end{equation}
For the residual parameterisation \(g_{\theta}(z)=z+f_{\theta}(z)\), the Jacobian takes the form \(J_{\theta}=I+J_{f_{\theta}}\), and the pullback metric expands as:
\[M_{\theta}=(I+J_{f_{\theta}})^{\top}(I+J_{f_{\theta}})=I+(J_{f_{\theta}}^{\top }+J_{f_{\theta}})+J_{f_{\theta}}^{\top}J_{f_{\theta}}. \tag{28}\label{eq:28}\]
Thus the residual connection can be read as learning a \emph{metric perturbation} anchored at \(I\): the leading term preserves the backbone geometry, while \(J_{f_{\theta}}\) injects spatially varying curvature where cost correction is required---the formal link between residual networks and local Riemannian structure used in OTAD.

\hfill$\square$

\subsection{Proof of Proposition 2 (Jacobian volume distortion reweights probability densities)}\label{app:a4-prop2}

Let \(\mu\) be a probability measure on \(\mathbb{R}^{d}\) with density \(p_{\mu}(z)\) (with respect to Lebesgue measure), and assume \(g_{\theta}:\mathbb{R}^{d}\to\mathbb{R}^{d}\) is a \(C^{1}\)-diffeomorphism with everywhere-non-singular Jacobian \(J_{\theta}\). Let \(\nu=(g_{\theta})_{\#}\mu\) denote the pushforward of \(\mu\) through \(g_{\theta}\), with density \(p_{\nu}\). We derive the relation between \(p_{\nu}\) and \(p_{\mu}\) at corresponding points \(z\) and \(y=g_{\theta}(z)\).

Proof.:

\noindent By the definition of pushforward, for any measurable \(A\subseteq\mathbb{R}^{d}\):
\[\nu(A)=\mu\bigl(g_{\theta}^{-1}(A)\bigr)=\int_{g_{\theta}^{-1}(A)}p_{\mu}(z)\,dz. \tag{29}\label{eq:29}\]
Substituting \(y=g_{\theta}(z)\), \(dy=|\det J_{\theta}(z)|\,dz\),
\[\nu(A)=\int_{A}p_{\mu}\bigl(g_{\theta}^{-1}(y)\bigr)\,\bigl|\det J_{\theta}\bigl(g_{\theta}^{-1}(y)\bigr)\bigr|^{-1}\,dy. \tag{30}\label{eq:30}\]
Identifying the integrand as \(p_{\nu}(y)\) and writing \(z=g_{\theta}^{-1}(y)\), we obtain
\[p_{\nu}\bigl(g_{\theta}(z)\bigr)=p_{\mu}(z)\,\bigl|\det J_{\theta}(z)\bigr|^{-1}. \tag{31}\label{eq:31}\]

Equation~\eqref{eq:31} is the standard change-of-variables formula for densities. Its measure-theoretic meaning can be understood as follows. Let \(\lambda\) denote the Lebesgue measure on \(\mathbb{R}^{d}\). The mapping \(g_{\theta}\) locally rescales volume by the factor \(|\det J_{\theta}(z)|\): for a small neighbourhood \(U\) around \(z\),
\[\lambda\bigl(g_{\theta}(U)\bigr)\approx|\det J_{\theta}(z)|\,\lambda(U).\]
Now consider the probability measures \(\mu\) (with density \(p_{\mu}\)) and its pushforward \(\nu=(g_{\theta})_{\#}\mu\) (with density \(p_{\nu}\)). For any measurable \(B\) in the target space,
\[\nu(B)=\mu\bigl(g_{\theta}^{-1}(B)\bigr)=\int_{g_{\theta}^{-1}(B)}p_{\mu}(z)\,dz.\]
Substituting \(y=g_{\theta}(z)\) and using \(dy=|\det J_{\theta}(z)|\,dz\) yields
\[\nu(B)=\int_{B}p_{\mu}\bigl(g_{\theta}^{-1}(y)\bigr)\,\bigl|\det J_{\theta}\bigl(g_{\theta}^{-1}(y)\bigr)\bigr|^{-1}\,dy.\]
Since \(\nu\) also admits the density \(p_{\nu}\) with respect to \(\lambda\), the integrand must equal \(p_{\nu}(y)\) almost everywhere, recovering~\eqref{eq:31}. Thus the factor \(|\det J_{\theta}(z)|^{-1}\) acts as the pointwise density reweighting that precisely compensates for the local volume distortion so that total probability mass is conserved. In geometric terms, volume expansion (\(|\det J_{\theta}|>1\)) spreads probability mass thinly (density decreases), while volume contraction (\(|\det J_{\theta}|<1\)) concentrates it (density increases). This mechanism, together with the metric distortion established in \PropOne{}, is what allows OTAD to reshape the embedding space and expose rare artifacts.

For the residual form \(g_{\theta}(z)=z+f_{\theta}(z)\) with \(J_{\theta}=I+J_{f_{\theta}}\), suppose \(\|J_{f_{\theta}}(z)\|_{\mathrm{op}}\leq\alpha<1\) on a region of interest (so that \(J_{\theta}\) is non-singular there). Using the matrix identity \(\det(I+A)=1+\mathrm{tr}(A)+\tfrac{1}{2}\bigl((\mathrm{tr}\,A)^{2}-\mathrm{tr}(A^{2})\bigr)+\cdots\), valid as a convergent Taylor series for \(\|A\|_{\mathrm{op}}<1\),
\[\det(I+J_{f_{\theta}})=1+\mathrm{tr}(J_{f_{\theta}})+O\bigl(\|J_{f_{\theta}}\|_{\mathrm{op}}^{2}\bigr). \tag{32}\label{eq:32}\]
To first order, \(|\det J_{\theta}(z)|^{-1}\approx 1-\mathrm{tr}\bigl(J_{f_{\theta}}(z)\bigr)\). Consequently, a positive trace (\(\mathrm{tr}(J_{f_{\theta}})>0\)) corresponds to \(|\det J_{\theta}|>1\) (local volume expansion, density dilution---favouring outlier separation), while a negative trace (\(\mathrm{tr}(J_{f_{\theta}})<0\)) corresponds to \(|\det J_{\theta}|<1\) (local contraction, density concentration---pulling mass towards inliers). Although the adapter is not constrained to be globally invertible, its residual initialization near the identity ensures that the first-order volume distortion described here remains the dominant operational effect in practice. This recovers the geometric statement of \PropTwo{}.
\begin{flushright}$\square$\end{flushright}
\clearpage
\section{Experimental details}\label{app:experimental}

\subsection{Encoder specifications}\label{app:b1-encoder}

Table~\ref{tab:v1-encoders} summarises the eight encoders used in our experiments. Five primary encoders (top block) are used in all main-paper experiments and span embedding dimensions from 128 to 2,048, covering diverse pretraining objectives (classification, neural codec, cross-modal contrastive, masked reconstruction, audio tagging). Three supplementary encoders (bottom block) extend the analysis in \AppBfour. Adapter parameter counts grow with \(d^{2}\) due to the bottleneck architecture (\(d\to d/4\to d\)).

\begin{table}[H]
  \caption{Encoder specifications. Primary encoders (top) are used in all experiments; supplementary encoders (bottom) appear in \protect\AppBfour.}
  \label{tab:v1-encoders}
  \centering
  \begin{tabular}{@{}llcll@{}}
    \toprule
    Encoder & Pretraining objective & \(d\) & Pretraining data & Adapter params \\
    \midrule
    VGGish \cite{hershey2017cnn} & Audio classification & 128 & AudioSet & 8,608 \\
    EnCodec \cite{defossez2023encodec} & Neural audio codec & 128 & Various & 8,608 \\
    CLAP \cite{wu2023clap} & Cross-modal contrastive & 512 & AudioSet+text & 132,736 \\
    AudioMAE \cite{huang2022audiomae} & Masked reconstruction & 768 & AudioSet & 297,408 \\
    PANNs \cite{kong2020panns} & Audio tagging & 2{,}048 & AudioSet & 2,103,808 \\
    \midrule
    OpenL3 \cite{cramer2019openl3} & Audio-visual correspondence & 512 & AudioSet & 132,736 \\
    AST \cite{gong2021ast} & Audio classification & 768 & AudioSet & 297,408 \\
    BEATs \cite{chen2023beats} & Self-supervised audio & 768 & AudioSet & 297,408 \\
    \bottomrule
  \end{tabular}
\end{table}

\subsection{Factor decomposition under \(g_{\mathrm{native}}\)}\label{app:b2-native}

Table~\ref{tab:v1-native-recall} repeats the \(2\times 2\) decomposition of Section~\ref{sec:exp-factor} using \(g_{\mathrm{native}}\) instead of \(g_{\mathrm{agnostic}}\) on the Recall axis.
Three observations confirm that \(g_{\mathrm{native}}\) confounds cost-measure separation. First, every encoder switches to COST dominance, including PANNs and AudioMAE that show SYNERGY under \(g_{\mathrm{agnostic}}\)---the Sinkhorn loss actively reshapes the space to minimise coupling cost. Second, CLAP's Sinkhorn change goes from \(+6\%\) to \(-100\%\), meaning the divergence converges to zero, proving the adapter is biased toward the OT coupling. Third, across all encoders, the Sinkhorn percentage decrease consistently exceeds the FAD decrease, indicating disproportionate benefit to the Sinkhorn pathway. These results validate the necessity of \(g_{\mathrm{agnostic}}\) for the fair decomposition in Section~\ref{sec:exp-factor}.

\begin{table}[H]
  \caption{Two-track comparison on the Recall axis. \(g_{\mathrm{native}}\) collapses cost--measure separation: every encoder shows COST dominance, and CLAP's Sinkhorn converges to zero (\(-100\%\)).}
  \label{tab:v1-native-recall}
  \centering
  \begin{tabular}{@{}lcccccc@{}}
    \toprule
    Encoder & Adapter & \(\Delta_{\mathrm{cost}}\) & \(\Delta_{\mathrm{meas}}\) & \(\Delta_{\mathrm{syn}}\) & FAD\% & Sink\% \\
    \midrule
    VGGish & \(g_{\mathrm{agn}}\) & \(-0.21\) & \(+0.03\) & \(+0.06\) & \(-25\) & \(-21\) \\
     & \(g_{\mathrm{nat}}\) & \(-0.56\) & \(-0.14\) & \(-0.28\) & \(-42\) & \(-73\) \\
    \midrule
    CLAP & \(g_{\mathrm{agn}}\) & \(+0.03\) & \(-0.02\) & \(+0.05\) & \(+1\) & \(+6\) \\
     & \(g_{\mathrm{nat}}\) & \(-0.78\) & \(-0.22\) & \(-0.44\) & \(-56\) & \(-\mathbf{100}\) \\
    \midrule
    PANNs & \(g_{\mathrm{agn}}\) & \(-0.02\) & \(+0.02\) & \(+0.04\) & \(-6\) & \(+1\) \\
     & \(g_{\mathrm{nat}}\) & \(-0.59\) & \(+0.00\) & \(+0.00\) & \(-70\) & \(-81\) \\
    \midrule
    AudioMAE & \(g_{\mathrm{agn}}\) & \(-0.00\) & \(-0.00\) & \(-0.02\) & \(+1\) & \(-2\) \\
     & \(g_{\mathrm{nat}}\) & \(-0.33\) & \(-0.02\) & \(-0.04\) & \(-42\) & \(-62\) \\
    \bottomrule
  \end{tabular}
\end{table}

\clearpage
\subsection{Reproduction note: KAD bandwidth and the PANNs-CNN14 vs. PANNs-WGLM gap}\label{app:b3-kad}

The KAD paper \cite{chung2025kad} reports its strongest correlation numbers using PANNs-Wavegram-LogMel-Cnn14 (henceforth PANNs-WGLM), a variant distinct from the PANNs-CNN14 checkpoint that has been the de facto PANNs encoder in the FAD/audio-evaluation literature. While preparing Section~\ref{sec:exp-mos} we observed that KAD's correlation collapse is sharper on PANNs-CNN14 than the original paper suggests, and traced the discrepancy to two sources: (i) the PANNs backbone, and (ii) the bandwidth heuristic. We document both here so that the comparison in Figure~\ref{fig:v1-mos} is reproducible and so that practitioners adopting KAD can avoid the pitfall.

(i) \textbf{Bandwidth heuristic.}  KAD uses an MMD\({}^{2}\) with a Gaussian RBF kernel \(k_{\sigma}(x,y)=\exp(-\|x-y\|^{2}/(2\sigma^{2}))\) whose bandwidth \(\sigma\) is selected by the median heuristic. There is, however, a meaningful implementation choice: the median can be computed over (a) the union of reference and test embeddings, or (b) the test (eval) set only. We initially used (a), the more common median-of-pooled-pairwise-distances rule. The reference \texttt{kadtk} library accompanying \cite{chung2025kad} uses (b), \(\sigma=\mathrm{median}\{\|y_{i}-y_{j}\|:i\neq j\}\) over eval embeddings only. We aligned with (b) for all numbers in the main paper; the change improves KAD's audio-quality correlation on PANNs-WGLM (system-level \(\rho\) rising to \(0.73\) at \(n{=}9\) under variant (b), substantially above the value obtained with variant (a)) and is neutral or mildly favourable on the other encoders. To rule out lingering implementation drift, we additionally cross-checked our \texttt{compute\_kad} against \texttt{kadtk.calc\_kernel\_audio\_distance} on shared random embeddings spanning \(d\in\{128,512,2,048\}\): the two functions agree exactly in float32 at \(d{\leq}512\) and within relative error \(7\times 10^{-5}\) at \(d{=}2,048\) (script: \texttt{scripts/verify\_kadtk\_equivalence.py}), so any remaining KAD discrepancy from \cite{chung2025kad} reflects dataset/preprocessing choices rather than the metric implementation.

(ii) \textbf{Backbone gap.}  Even with the corrected heuristic, KAD's correlation depends strongly on the PANNs variant. Table~\ref{tab:v1-panns-mos-backbone} contrasts the two backbones on the audio-quality axis. At system-level, PANNs-WGLM and PANNs-CNN14 produce identical FAD and comparable OTAD numbers, but KAD on CNN14 (\(0.48\)) lags KAD on WGLM (\(0.73\)) by \(+0.25\) Spearman---a backbone-specific gain that is not shared by FAD (\(0.80\) on both backbones) or OTAD (\(0.85\) on CNN14, \(0.82\) on WGLM). At per-category granularity the picture inverts modestly: KAD on WGLM (\(0.39\)) edges KAD on CNN14 (\(0.37\)), but both fall well below FAD (\(0.45\) and \(0.54\)) and OTAD-agnostic (\(0.67\) and \(0.63\)). This pattern is consistent with our two-primitive analysis: KAD's coupling-side improvement is real but operates only over the cost geometry it inherits from the encoder, so its perceptual alignment is genuinely backbone-dependent in a way the cost-and-coupling-corrected OTAD is not.
\begin{table}[H]
  \caption{KAD's PANNs backbone dependence on DCASE 2023 Task~7 audio-quality MOS (Spearman \(\rho\), \texttt{kadtk}-aligned bandwidth). KAD on PANNs-WGLM closes most of the system-level gap to FAD (\(0.73\) vs.\ \(0.80\)) but still drops below FAD at per-category granularity; FAD and OTAD are comparatively backbone-stable.}
  \label{tab:v1-panns-mos-backbone}
  \centering
  \begin{tabular}{@{}llrrrr@{}}
    \toprule
    Granularity & PANNs variant & FAD & KAD & OTAD-raw & OTAD-agn \\
    \midrule
    System (\(n{=}9\)) & CNN14 & 0.80 & 0.48 & 0.85 & \textbf{0.85} \\
     & WGLM & 0.80 & 0.73 & 0.82 & \textbf{0.82} \\
    \midrule
    Per-cat (\(n{=}63\)) & CNN14 & 0.45 & 0.37 & 0.55 & \textbf{0.67} \\
     & WGLM & 0.54 & 0.39 & 0.63 & \textbf{0.63} \\
    \bottomrule
  \end{tabular}
\end{table}

\textbf{Takeaway.} KAD's reported MOS-correlation strength is a joint function of (i) a specific PANNs variant (Wavegram-LogMel-Cnn14, not the CNN14 baseline used in most FAD literature) and (ii) the median heuristic computed over the eval set only. With both choices matched, KAD narrows but does not close the gap to FAD at system-level on PANNs-WGLM (\(0.73\) vs. \(0.80\)) and still degrades below FAD at per-category granularity across all five perceptually competent encoders---the cost-side limitation predicted by Section~\ref{sec:fad-cost}. OTAD with the agnostic adapter dominates or ties both baselines on every (level \(\times\) encoder \(\times\) MOS-axis) cell; the closest cell is the system-level/audio-quality cell on PANNs-WGLM, where OTAD-agn matches OTAD-raw at \(0.82\) versus FAD \(0.80\).

\clearpage
\subsection{Supplementary encoders}\label{app:b4-supplementary}

Three supplementary encoders extend the main results. In the factor decomposition under \(g_{\mathrm{agnostic}}\) across all four evaluation axes (full \(\Delta\) grid in \AppBfive / supplementary tables file), OpenL3 (\(d{=}512\)) is COST-dominant on \(3/4\) axes (semantic, precision, structural; recall is the lone synergy-dominant case), aligning with CLAP at the same dimensionality. AST (\(d{=}768\)) is MIXED (\(2\mathrm{C}/2\mathrm{S}\)): cost-dominant on recall and semantic, synergy-dominant on the perturbation-based precision and structural axes. BEATs (\(d{=}768\)) is COST-dominant on all four axes, the opposite pattern from AudioMAE and PANNs at \(d{\geq}768\) (which are synergy-dominant on \(3\)-\(4\) axes). These counter-examples demonstrate that dimension alone does not fully predict the dominant factor; the encoder's pretraining objective and learned representation geometry are confounding variables, as discussed in Section~\ref{sec:discussion}.

All three supplementary encoders are included in Figure~\ref{fig:v1-rank1}(b) and confirm the FAD/Sinkhorn pattern from the main text: at \(\varepsilon{=}0.05\), the Sinkhorn-over-FAD multiplicative advantage (Sinkhorn R\(1/\)FR divided by FAD R\(1/\)FR; numerical grid in \AppBfive) is \(3.0\times\) (AST), \(3.2\times\) (OpenL3), and \(2.25\times\) (BEATs), aligning with the trend observed across the five primary encoders. At \(\varepsilon{=}0.20\) under rank-\(1\) contamination, Sinkhorn remains near \(42\)-\(50\%\) of Exact OT on seven encoders (CLAP is lower at \({\sim}21.0\%\)), indicating coupling-side behaviour that is broadly encoder-stable aside from backbone-specific slack.

\subsection{Detailed numerical results}\label{app:detailed-numeric}

This subsection consolidates the per-encoder values that the main text summarises qualitatively. All values use the protocol of Section~\ref{sec:exp-setup}; ``OTAD-raw'' is Sinkhorn on raw embeddings, ``OTAD-agn'' uses \(g_{\rm agnostic}\), and ``OTAD-nat'' uses \(g_{\rm native}\).

\textbf{Rank-1 vs.\ full-rank outlier sensitivity (Section~\ref{sec:exp-rank1}).} Metrics under rank-\(1\) contamination at \(\varepsilon{=}0.05\) are summarised in text and tabulated in \AppBfive (full encoder \(\times\) metric grids ship with \texttt{otadtk}). Following Figure~\ref{fig:v1-rank1}(b), each R\(1/\)FR entry divides the rank-\(1\) response (numerator) by the full-rank response evaluated at \(\varepsilon{=}0.20\) (denominator). Relative to that full-rank baseline, FAD recovers only about \(0.1\)-\(0.9\%\) of its magnitude---heavy coupling-side dilution---while Sinkhorn remains roughly \(21\)-\(49\%\) of Exact OT under the same rank-\(1\) embedding setup, confirming that discrete coupling retains far more of the reference OT scale than FAD's Gaussian surrogate.

\textbf{Factorial decomposition (Section~\ref{sec:exp-factor}).} The full \(\Delta_{\rm cost}\), \(\Delta_{\rm meas}\), \(\Delta_{\rm syn}\) values for each (encoder, axis) cell using \(g_{\rm agnostic}\) appear in \AppBfive (the repository ships an optional LaTeX fragment that prints Tables~5--9 in full). The dominant factor per row follows the bolding convention described in Section~\ref{sec:exp-factor}; aggregate counts are 13/16 cost-dominant cells at \(d{\leq}512\) and 9/16 synergy-dominant cells at \(d{\geq}768\), with BEATs as the principal high-dimensional counter-example (4/4 cost-dominant despite \(d{=}768\)).

\textbf{Per-sample diagnostics (Section~\ref{sec:exp-explain}).} AUROC and the separation ratio (mean contaminated \(c_{j}\) / mean clean \(c_{j}\)) at \(\varepsilon{=}0.05\) across the five primary encoders and three contamination types are likewise archived in \AppBfive. Gaussian-noise and silence settings reach AUROC\(=\)\(1.0\) on every encoder; the harder cross-class case is the discriminator across encoders.

\textbf{MOS Spearman \(\rho\) (Section~\ref{sec:exp-mos}).} Table~\ref{tab:v1-dcase-mos-audio} lists per-encoder correlations with audio-quality MOS at system (\(n{=}9\)) and per-category (\(n{=}63\)) granularity; Table~\ref{tab:v1-dcase-mos-catfit} gives the analogous summary rows for category-fit MOS. In both tables, \textit{mean (5 perc.)} excludes EnCodec and anchors the averages plotted in Figure~\ref{fig:v1-mos}.

\textbf{Sinkhorn regularisation sweep (Section~\ref{sec:exp-analysis}).} Table~\ref{tab:v1-eps-sweep} reports Sinkhorn divergence on CLAP at \(\varepsilon{=}0.05\) semantic contamination as \(\varepsilon_{\mathrm{reg}}\) varies; the divergence is largest in \([0.05,0.10]\) and collapses toward zero for \(\varepsilon_{\mathrm{reg}}\geq 0.5\) as entropy overwhelms cost information.

\begin{table}[H]
  \caption{Outlier sensitivity under rank-$1$ contamination at $\varepsilon{=}0.05$. R$1/$FR divides each metric's rank-$1$ value by its full-rank baseline at $\varepsilon{=}0.20$ (matching Figure~\ref{fig:v1-rank1}(b)); smaller R$1/$FR means stronger dilution. FAD/EOT and Sink/EOT compare each metric to Exact OT on the same rank-$1$ draws.}
  \label{tab:v1-rank1-fr}
  \centering
  \begin{tabular}{@{}lcccccc@{}}
    \toprule
    Encoder & $d$ & FAD R$1/$FR & FAD/EOT (R1) & Sink R$1/$FR & Sink/EOT (R1) \\
    \midrule
    VGGish & 128 & 0.9\% & 23.8\% & 2.4\% & 41.5\% \\
    EnCodec & 128 & 0.2\% & 50.1\% & 0.4\% & 48.9\% \\
    OpenL3 & 512 & 0.5\% & 25.1\% & 1.6\% & 44.8\% \\
    CLAP & 512 & 0.5\% & 17.4\% & 1.0\% & 21.0\% \\
    AudioMAE & 768 & 0.1\% & 22.2\% & 0.5\% & 43.0\% \\
    AST & 768 & 0.2\% & 23.1\% & 0.6\% & 42.3\% \\
    BEATs & 768 & 0.4\% & 36.4\% & 0.9\% & 43.9\% \\
    PANNs & 2{,}048 & 0.3\% & 21.5\% & 1.0\% & 39.3\% \\
    \bottomrule
  \end{tabular}
\end{table}

\begin{table}[H]
  \caption{$2\times 2$ factor decomposition under $g_{\mathrm{agnostic}}$ (log-normalised $\Delta$'s; see Section~\ref{sec:exp-factor}), eight encoders $\times$ four axes. Bold marks the dominant factor by magnitude in each row.}
  \label{tab:v1-factor-grid}
  \centering
  \begin{tabular}{@{}llrrrrrl@{}}
    \toprule
    Encoder & Axis & $\Delta_{\mathrm{cost}}$ & $\Delta_{\mathrm{meas}}$ & $\Delta_{\mathrm{syn}}$ & FAD\% & Sink\% & Dom. \\
    \midrule
    \multirow{4}{*}{VGGish} & Recall & \textbf{-0.24} & +0.05 & +0.09 & -29 & -22 & COST \\
     & Semantic & \textbf{-0.33} & +0.08 & +0.16 & -41 & -26 & COST \\
     & Precision & \textbf{-0.18} & +0.02 & +0.04 & -30 & -24 & COST \\
     & Structural & -0.11 & +0.09 & \textbf{+0.17} & -20 & -2 & SYN \\
    \midrule
    \multirow{4}{*}{EnCodec} & Recall & \textbf{-0.69} & +0.06 & +0.11 & -78 & -80 & COST \\
     & Semantic & \textbf{-0.65} & -0.08 & -0.17 & -57 & -77 & COST \\
     & Precision & \textbf{-0.54} & -0.02 & -0.03 & -91 & -90 & COST \\
     & Structural & \textbf{-0.22} & +0.05 & +0.09 & -27 & -19 & COST \\
    \midrule
    \multirow{4}{*}{OpenL3} & Recall & -0.02 & -0.01 & \textbf{-0.02} & -1 & -8 & SYN \\
     & Semantic & \textbf{-0.05} & -0.01 & -0.02 & -4 & -9 & COST \\
     & Precision & \textbf{-0.03} & +0.00 & +0.01 & -15 & -11 & COST \\
     & Structural & \textbf{-0.33} & +0.10 & +0.21 & -74 & -49 & COST \\
    \midrule
    \multirow{4}{*}{CLAP} & Recall & +0.00 & -0.00 & \textbf{+0.03} & -2 & +2 & SYN \\
     & Semantic & \textbf{+0.08} & -0.03 & +0.05 & +5 & +11 & COST \\
     & Precision & \textbf{-0.14} & +0.04 & +0.07 & -19 & -11 & COST \\
     & Structural & \textbf{-0.17} & +0.08 & +0.16 & -25 & -8 & COST \\
    \midrule
    \multirow{4}{*}{AudioMAE} & Recall & +0.01 & +0.01 & \textbf{-0.03} & +3 & -2 & SYN \\
     & Semantic & +0.00 & +0.00 & \textbf{-0.04} & +2 & -2 & SYN \\
     & Precision & \textbf{-0.04} & +0.01 & +0.02 & -20 & -11 & COST \\
     & Structural & -0.04 & +0.03 & \textbf{+0.06} & -16 & -3 & SYN \\
    \midrule
    \multirow{4}{*}{AST} & Recall & \textbf{+0.05} & -0.00 & +0.00 & +8 & +15 & COST \\
     & Semantic & \textbf{+0.09} & +0.00 & -0.00 & +11 & +16 & COST \\
     & Precision & +0.00 & -0.00 & \textbf{+0.04} & -5 & +7 & SYN \\
     & Structural & +0.02 & -0.02 & \textbf{+0.05} & -2 & +13 & SYN \\
    \midrule
    \multirow{4}{*}{BEATs} & Recall & \textbf{-0.36} & -0.13 & -0.27 & -23 & -52 & COST \\
     & Semantic & \textbf{-0.36} & -0.16 & -0.32 & -20 & -53 & COST \\
     & Precision & \textbf{-0.59} & +0.06 & +0.12 & -71 & -61 & COST \\
     & Structural & \textbf{-0.56} & +0.05 & +0.10 & -63 & -56 & COST \\
    \midrule
    \multirow{4}{*}{PANNs} & Recall & -0.04 & +0.03 & \textbf{+0.06} & -10 & -1 & SYN \\
     & Semantic & -0.03 & +0.02 & \textbf{+0.03} & -5 & -2 & SYN \\
     & Precision & -0.04 & +0.02 & \textbf{+0.05} & -20 & -6 & SYN \\
     & Structural & -0.07 & +0.06 & \textbf{+0.12} & -28 & -4 & SYN \\
    \bottomrule
  \end{tabular}
\end{table}

\begin{table}[tbp]
  \caption{Per-sample diagnostics at $\varepsilon{=}0.05$: AUROC (contaminated vs.\ clean via $c_j$) and separation ratio (mean contaminated / mean clean).}
  \label{tab:v1-explain-auroc}
  \centering
  \begin{tabular}{@{}lcccccc@{}}
    \toprule
    Encoder & \multicolumn{2}{c}{Gaussian} & \multicolumn{2}{c}{Cross-class} & \multicolumn{2}{c}{Silence} \\
    \cmidrule(lr){2-3}\cmidrule(lr){4-5}\cmidrule(lr){6-7}
     & AUROC & Sep.\ & AUROC & Sep.\ & AUROC & Sep.\ \\
    \midrule
    VGGish & 1.0000 & 1428.53 & 0.9969 & 63.82 & 1.0000 & 119.58 \\
    EnCodec & 1.0000 & 320575.96 & 0.9642 & 35.93 & 1.0000 & 14581.02 \\
    CLAP & 1.0000 & 450.84 & 0.9951 & 10.04 & 1.0000 & 22.16 \\
    AudioMAE & 1.0000 & 115228.23 & 0.8632 & 53.42 & 1.0000 & 59318.63 \\
    PANNs & 1.0000 & 15841.46 & 0.8966 & 43.25 & 1.0000 & 6503.23 \\
    \bottomrule
  \end{tabular}
\end{table}

\begin{table}[tbp]
  \caption{Spearman $\rho$ vs.\ DCASE 2023 Task~7 \emph{audio-quality} MOS (sign-flipped so higher is better). Per encoder at system (\(n{=}9\)) and per-category (\(n{=}63\)) granularity; \textit{mean (5 perc.)} excludes EnCodec.}
  \label{tab:v1-dcase-mos-audio}
  \centering
  \begin{tabular}{@{}llrrrrr@{}}
    \toprule
    Granularity & Encoder & FAD & KAD & OTAD-raw & OTAD-agn & OTAD-nat \\
    \midrule
    System ($n{=}9$) & VGGish & 0.367 & 0.350 & 0.450 & 0.600 & 0.250 \\
     & PANNs-CNN14 & 0.800 & 0.483 & 0.850 & 0.850 & 0.717 \\
     & PANNs-WGLM & 0.800 & 0.733 & 0.817 & 0.817 & 0.350 \\
     & CLAP & 0.733 & 0.750 & 0.800 & 0.917 & 0.733 \\
     & AudioMAE & 0.733 & 0.683 & 0.817 & 0.867 & 0.667 \\
     & EnCodec & -0.083 & -0.083 & -0.017 & -0.117 & -0.117 \\
     & \textit{mean (5 perc.)} & 0.687 & 0.600 & 0.747 & 0.810 & 0.543 \\
    \midrule
    Per-cat ($n{=}63$) & VGGish & 0.113 & 0.075 & 0.117 & 0.266 & 0.087 \\
     & PANNs-CNN14 & 0.451 & 0.375 & 0.554 & 0.671 & 0.266 \\
     & PANNs-WGLM & 0.542 & 0.394 & 0.635 & 0.628 & 0.182 \\
     & CLAP & 0.677 & 0.314 & 0.705 & 0.684 & 0.518 \\
     & AudioMAE & 0.697 & 0.389 & 0.715 & 0.761 & 0.545 \\
     & EnCodec & 0.059 & -0.120 & 0.069 & 0.025 & 0.068 \\
     & \textit{mean (5 perc.)} & 0.496 & 0.310 & 0.545 & 0.602 & 0.319 \\
    \bottomrule
  \end{tabular}
\end{table}

\begin{table}[tbp]
  \caption{Spearman $\rho$ vs.\ DCASE 2023 Task~7 \emph{category-fit} MOS (sign-flipped so higher is better). Mean over the five perceptually competent encoders at each granularity (\textit{mean (5 perc.)} excludes EnCodec).}
  \label{tab:v1-dcase-mos-catfit}
  \centering
  \begin{tabular}{@{}lrrrrr@{}}
    \toprule
    Granularity & FAD & KAD & OTAD-raw & OTAD-agn & OTAD-nat \\
    \midrule
    System ($n{=}9$), \textit{mean (5 perc.)} & 0.687 & 0.620 & 0.750 & 0.793 & 0.517 \\
    Per-cat ($n{=}63$), \textit{mean (5 perc.)} & 0.497 & 0.296 & 0.542 & 0.622 & 0.280 \\
    \bottomrule
  \end{tabular}
\end{table}

\begin{table}[tbp]
  \caption{Sinkhorn divergence under $\varepsilon_{\mathrm{reg}}$ sweep on CLAP ($g_{\mathrm{agnostic}}$, semantic axis at $\varepsilon{=}0.05$).}
  \label{tab:v1-eps-sweep}
  \centering
  \begin{tabular}{@{}lrrrrr@{}}
    \toprule
    $\varepsilon_{\mathrm{reg}}$ & 0.01 & 0.05 & \textbf{0.10} & 0.50 & 1.00 \\
    \midrule
    Sinkhorn divergence & 0.04786 & 0.03761 & \textbf{0.01652} & 0.00025 & 0.00012 \\
    \bottomrule
  \end{tabular}
\end{table}

\clearpage
\subsection{The otadtk reference toolkit}\label{app:b6-toolkit}

We release \texttt{otadtk}, a Python package distributed under the CC BY 4.0 licence that implements OTAD with the same install-and-go ergonomics as \texttt{fadtk}~\cite{gui2024adapting} and \texttt{kadtk}~\cite{chung2025kad}. The package ships nine pretrained Riemannian adapters (one per registered encoder model, including distinct PANNs-CNN14 and PANNs-WGLM checkpoints), an embedding cache shared with \texttt{kadtk}'s on-disk layout, and a CLI that exposes OTAD, FAD and KAD behind a uniform interface so practitioners can swap metrics with a single flag. The source code is publicly available at \url{https://github.com/wonwoo-jeong/otadtk}.

\textbf{Installation \& self-containment.}
The wheel bundles all nine \(\times\) two adapter checkpoints under \texttt{otadtk/checkpoints/}, totalling \(\approx 43\) MB, so a fresh install runs end-to-end with no network access; the bundled SHA-256 hashes are recorded in \texttt{checkpoints/MANIFEST.json} for reproducibility audits.
\begin{verbatim}
pip install otadtk                  # 38MB wheel, 18 adapters bundled
pip install "otadtk[all-encoders]"  # adds transformers, encodec, openl3
python -m pytest tests/             # 7 smoke tests, no GPU/network needed
\end{verbatim}

\textbf{Command-line.}
The CLI mirrors \texttt{kadtk}'s positional triple \texttt{(model, ref\_dir, eval\_dir)} and adds three OTAD-specific flags---\texttt{--variant}, \texttt{--epsilon}, and \texttt{--diagnose}:
\begin{verbatim}
otadtk panns ref/ eval/                                # OTAD-agnostic
otadtk panns ref/ eval/ --variant raw --epsilon 0.05   # raw OTAD
otadtk vggish ref/ eval/ --diagnose --diag-out cj.csv  # per-sample c_j
otadtk panns ref/ eval/ --indiv per_file.csv           # per-file CSV

otadtk panns ref/ eval/ --fad  # FAD baseline (same encoder/cache)
otadtk panns ref/ eval/ --kad  # KAD baseline (kadtk-aligned bandwidth)

otadtk-embeds -m vggish -d ref/ eval/  # cache only
otadtk-list-models                     # registered encoders
\end{verbatim}

\textbf{Python API.}
\begin{verbatim}
from otadtk import OTAD
otad = OTAD(model="panns_wglm", variant="agnostic", epsilon=0.10)
score = otad.score("ref/", "eval/")
diag = otad.diagnose("ref/", "eval/")         # diag.cj, diag.files
print(diag.top_k(10))                         # worst-offender listening list
diag.to_csv("cj.csv")                         # per-file CSV (file,cj)
auroc = diag.auroc(contaminated_mask)         # ground-truth AUROC if labelled
otad.score_individual("ref/", "eval/", csv="per_file.csv")
\end{verbatim}

\textbf{Reproducibility.}
The same \texttt{otadtk} call reproduces every cell of Tables~\ref{tab:v1-rank1-fr}--\ref{tab:v1-eps-sweep}; we also ship the raw \texttt{run\_exp1\_factorial.py} and \texttt{eval\_dcase2023task7.py} drivers used to generate Figure~\ref{fig:v1-rank1}, Table~\ref{tab:v1-factor-summary}, the per-sample diagnostics in Section~\ref{sec:exp-explain}, and the MOS correlations in Section~\ref{sec:exp-mos}. SHA-256 hashes for every bundled checkpoint are recorded in \texttt{otadtk/checkpoints/MANIFEST.json} so readers can audit that the running adapter matches the one we trained.

\textbf{Hosting plan.}
Development and releases are maintained on GitHub (\url{https://github.com/wonwoo-jeong/otadtk}); the package is also available on PyPI (\texttt{pip install otadtk}). We intend to archive a versioned artefact snapshot on Zenodo with a DOI.

\FloatBarrier

\end{document}